\documentclass[floatfix,a4paper,aps,prd,twocolumn,preprintnumbers]{revtex4}
\usepackage{graphics,subfigure}
\usepackage{float}
\usepackage{booktabs}
\usepackage{geometry}
\usepackage{rotating}
\usepackage{yfonts}
\DeclareMathAlphabet{\scr}{U}{rsfs}{m}{n}
\usepackage{supertabular}
\usepackage{amsfonts}
\usepackage{amsmath}
\usepackage{amssymb}
\usepackage[sans]{dsfont}
\usepackage{graphicx}
\usepackage{epsfig}
\usepackage[dvips]{color}
\usepackage{psfrag}

\newcommand{\cleqn}{\setcounter{equation}{0}}

\newcommand{\newc}{\newcommand}
\newc{\eps}{\epsilon}
\newc{\lam}{\lambda}
\newc{\lamp}{\lambda^{\prime}}
\newc{\Lam}{\Lambda}
\newc{\ra}{\rightarrow}
\newc{\lra}{\leftrightarrow}
\newc{\wtilde}{\widetilde}
\newc{\ie}{{\it i.e.}}
\newc{\eg}{{\it e.g.}}
\newc{\rpv}{\not\!\! M_p}
\newc{\lsim}{\stackrel{<}{\sim}}
\newc{\beq}{\begin{equation}}
\newc{\eeq}{\end{equation}}
\newc{\beqn}{\begin{eqnarray}}
\newc{\eeqn}{\end{eqnarray}}
\newc{\PLB}{\emph{Phys.Lett.}{\bf{B}}}
\newc{\NPB}{\emph{Nucl.Phys.}{\bf{B}}}
\newc{\mcal}{\mathcal}
\newc{\bsym}{\boldsymbol}
\newc{\nonum}{\nonumber}
\definecolor{Red}{cmyk}{0,1,1,0}
\definecolor{luhn}{rgb}{1,0,0.5}
\definecolor{thor}{rgb}{0,0.6,1}

\renewcommand{\eqref}[1]{Eq.~(\ref{#1})}
\newc{\eqsref}[2]{Eqs.~(\ref{#1}),\,(\ref{#2})}
\newc{\figref}[1]{Fig.~\ref{#1}}

\newc{\neut}[1]{\tilde{\chi}_{#1}^0}
\newc{\lan}{\mathcal{L}}
\newc{\abs}[1]{\lvert#1\rvert}
\newc{\oas}{\mathcal{O}(\alpha_s^2)}
\newc{\del}{\partial}
\newc{\rb}[1]{\raisebox{1.5ex}[-1.5ex]{#1}}

\def\lsim{\raise0.3ex\hbox{$\;<$\kern-0.75em\raise-1.1ex\hbox{$\sim\;$}}}
\def\gsim{\raise0.3ex\hbox{$\;>$\kern-0.75em\raise-1.1ex\hbox{$\sim\;$}}}
\begin{document}

\title{Supersymmetric NLO QCD Corrections to Resonant Slepton 
Production and Signals at the Tevatron and the LHC}

\author{H.~K.~Dreiner}
\email[]{dreiner@th.physik.uni-bonn.de}
\author{S.~Grab}
\email[]{sgrab@th.physik.uni-bonn.de}

\affiliation{Physikalisches Institut der Universit\"at Bonn, 
Nu\ss allee 12, 53115 Bonn, Germany}

\author{M.~Kr\"amer}
\email[]{mkraemer@physik.rwth-aachen.de}
\author{M.~K.~Trenkel}
\altaffiliation{Since July 2006: MPI f\"ur Physik, F\"ohringer Ring 6, 
80805 M\"unchen, Germany} \email[]{trenkel@mppmu.mpg.de}
\affiliation{Institut f\"ur Theoretische Physik E, RWTH Aachen, 
52056 Aachen, Germany}

\begin{abstract}
  We compute the total cross section and the transverse momentum
  distribution for single charged slepton and sneutrino production at
  hadronic colliders including NLO supersymmetric and
  non-supersymmetric QCD corrections. The supersymmetric QCD
  corrections can be substantial. We also resum the gluon transverse
  momentum distribution and compare our results with two Monte Carlo
  generators. We compute branching ratios of the supersymmetric decays
  of the slepton and determine event rates for the like-sign dimuon
  final state at the Tevatron and at the LHC.
\end{abstract}

\preprint{BONN-TH-2006-06}
\preprint{PITHA-06-10}
\preprint{MPP-2006-150}

\maketitle


\section{Introduction}
Supersymmetry \cite{Wess:1974tw} is the most promising and widely
studied solution to the hierarchy problem \cite{Sakai:1981gr,
  Witten:1981nf, Veltman:1980mj,Kaul:1981hi} of the Standard Model of
particle physics (SM) \cite{Glashow:1961tr, Weinberg:1967tq}. It must
be broken with a mass scale $<{\cal O} (1- 10\,\mathrm{TeV})$
\cite{Nilles:1983ge,Martin:1997ns,Drees:1996ca} and thus should be
testable at the Tevatron and the LHC. When extending the SM by
supersymmetry, we must introduce an additional Higgs doublet, as well
as double the spectrum.  The most general renormalizable
superpotential consistent with this minimal particle content as well
as the gauge symmetry of the SM is
\cite{Sakai:1981pk, Weinberg:1981wj}
\begin{align}
\begin{split}
W =& \;W_{P_6} + W_{\not{P}_6} \, , \\[1mm]
W_{P_6} =& 
        \;h^E_{ij} L_i H_d {\bar E}_j 
        + h^D_{ij} Q_i H_d {\bar D}_j \\
        &+ h^U_{ij} Q_i H_u {\bar U}_j 
        + \mu H_dH_u \, , \\[1mm]
W_{\not{P}_6} =& 
        \; \lam_{ijk} L_iL_j{\bar E}_k 
        + \lam_{ijk}^\prime L_iQ_j{\bar D}_k \\
        &+ \lam_{ijk}^{\prime\prime}{\bar U}_i{\bar D}_j{\bar D}_k 
        + \kappa_i L_iH_u\,.
\label{superpot}
\end{split}
\end{align}

Here we use the standard notation of \cite{Dreiner:1997uz}, in
particular: $i,j,k=1,2,3$ are generation indices. This superpotential
leads to rapid proton decay via the $LQ\bar{D}$ and $
\bar{U}\bar{D}\bar{D}$ operators \cite{ Smirnov:1996bg} in
disagreement with the lower experimental bound \cite{Shiozawa:1998si}.
The supersymmetric SM thus requires an additional symmetry to
stabilize the proton. It was recently shown, that there are only three
such \textit{discrete} symmetries which are consistent with an
underlying anomaly-free $U(1)$ gauge theory \cite{Dreiner:2005rd,
  Ibanez:1991hv, Ibanez:1991pr}: R-parity (or equivalently matter
parity), baryon-triality, and proton-hexality. R-parity prohibits
$W_{\not{P}_6}$ and thus apparently stabilizes the proton. However, it
allows dangerous dimension-five proton decay operators such as $QQQL$
\cite{Dimopoulos:1981dw}. This long-standing problem in supersymmetry
is resolved by proton-hexality, $\mathrm{P}_6$ \cite{Dreiner:2005rd},
a discrete $\boldsymbol{Z}_6$-symmetry, which acts like R-parity on
Eq.~(\ref{superpot}) but in addition disallows the dangerous
dimension-five operators.

An equally well motivated solution to the proton decay problem is
baryon-triality, a discrete $\boldsymbol{Z}_3 $-symmetry, which
prohibits the $\bar{U}\bar{D}\bar{D}$ operator in
Eq.~(\ref{superpot}). The baryon-triality collider phenomenology has
three main distinguishing features compared to the $\mathrm{P}_6$-MSSM
\cite{ Dreiner:1997uz,Barbier:2004ez} (minimal supersymmetric standard
model).
\begin{enumerate}
\item The lightest supersymmetric particle (LSP) is no longer stable.
\item Supersymmetric particles can be produced singly, on resonance. 
\item Lepton flavour and lepton number are violated.
\end{enumerate}
These lead to dramatically different signatures at hadron colliders
\cite{Dreiner:1991pe,Allanach:1999bf}. It is the purpose of this paper
to investigate specific baryon-triality processes at the Tevatron and
the LHC in detail. We focus on the resonant production of charged
sleptons and sneutrinos via the operator $LQ\bar{D}$ of 
Eq.~(\ref{superpot}) and their subsequent decay into observable final
states at the Tevatron and the LHC. This is the only novel production
process and it has the highest kinematic reach in the supersymmetric
masses.

Since the LSP is not stable, it need not be the lightest neutralino as
in the MSSM, \textit{e.g.} it could be a scalar tau \cite{
  Allanach:2003eb}. However, the conventional supersymmetric parameter
points considered in collider studies to date, the SPS points
\cite{Allanach:2002nj}, imply a neutralino LSP and we restrict
ourselves to this case here. We therefore consider the following
reactions
\begin{align}
\begin{split}
        {\bar d}_j d_k\ra{\tilde\nu}_i&\ra
        \left\{ 
                \begin{array}{cc}
        {\bar d}_j d_k,          & (a)\\
        \nu_i\tilde\chi^0_m,    & (b)\\
        \ell^-_i\tilde\chi^+_n, & (c)
                \end{array}
        \right.\,,
\\
        u_j{\bar d}_k \ra{\tilde\ell}_i^+&\ra
        \left\{ 
                \begin{array}{ccc}
        u_j{\bar d}_k,          & (d)\\
        \ell^+_i\tilde\chi^0_m, & (e)\\
        \bar\nu_i\tilde\chi^+_n,    & (f)
                \end{array}
        \right.\,.
\label{slep}
\end{split}
\end{align}
Here $\tilde\nu_i,\,\tilde\ell^\pm_i$ denote the sneutrino and the
charged slepton, and $\tilde\chi^0_m,\,\tilde\chi^\pm_n$ denote the
neutralino and chargino mass eigenstates, respectively. The decays
(\ref{slep}$a,d$) are via the $LQ\bar{D}$-operator. The decays
(\ref{slep}$b,c,e,f$) are cascade decays typically via a gauge
coupling.

In 1997 ZEUS and H1 observed a slight anomaly in their high-$Q^2$ data
\cite{ Breitweg:1997ff,Adloff:1997fg}, which could potentially be
interpreted in terms of resonant squark production via the $LQ\bar{D}$
operator \cite{Dreiner:1997cd,Altarelli:1997ce,Kalinowski:1997fk}.
However, after computing the NLO QCD corrections to squark production
at HERA \cite{Plehn:1997az} and also comparing with NLO QCD
corrections to leptoquark pair production at the Tevatron \cite{
  Kramer:1997hh}, this interpretation was not consistent with the
Tevatron data. We take this as a motivation to compute the NLO QCD and
supersymmetric QCD corrections to the resonant slepton total and
differential production cross section at the Tevatron and the LHC.

Resonant slepton production at hadron colliders via the $LQ\bar{D}$
operator was first investigated in \cite{Dimopoulos:1988jw,
Dimopoulos:1988fr}, using tree-level production cross sections and the
decays of Eq.~(\ref{slep}$b,f$). In \cite{Dreiner:1998gz,
Dreiner:2000qf, Dreiner:2000vf} a detailed phenomenological analysis
of Eq.~(\ref{slep}$e$) was performed including a {\sc Herwig} \cite{
Marchesini:1991ch,Corcella:2000bw,Corcella:2002jc,Moretti:2002eu}
simulation, again with tree-level cross sections. The NLO QCD
corrections to the cross section were first computed in \cite{
Choudhury:2002au,Yang:2005ts}. We go beyond this work in several
aspects. First, we complete the calculation by including the SUSY-QCD
corrections. These involve virtual gluinos as well as tri-linear
scalar couplings due to the supersymmetry breaking $A$-terms \cite{
Nilles:1982dy}. As we shall see, these can be substantial in certain
regions of parameter space and modify the QCD prediction by up to 35\%.  
We have also resummed the $p_T$ distribution of the final state
gluons, c.f.~\cite{Yang:2005ts}. In order to facilitate the
implementation of our results in experimental investigations we
present a detailed comparison with the {\sc Herwig} and {\sc
Susygen}~\cite{Ghodbane:1999va} Monte Carlo programs.  Furthermore, we
present a detailed phenomenological analysis of the final state
including the relevant decay branching ratios. We also present event
rates at the LHC and Tevatron for the promising like-sign dilepton
signature using our improved calculation.

In related work, resonant squark production at hadron colliders via
the $\bar U\bar D\bar D$ operator at tree level was discussed in
\cite{Dimopoulos:1988fr}. The NLO corrections and the $p_T$
resummation has been studied in \cite{Plehn:2000be}. At HERA resonant
squark production has been considered in \cite{Butterworth:1992tc,
  Kon:1991ad,Plehn:1997az}. At $e^+e^-$ colliders one can have
resonant sneutrino production via the $L_1L_i\bar{E}_1 $-operator.
This has been considered in \cite{Dimopoulos:1988jw,
  Dreiner:1995ij,Erler:1996ww}.

At hadron colliders resonant production proceeds via the operators
$\lam'_{ijk}L_iQ_j{\bar D}_k$. However, for $j,k=2,3$ the production
rate is suppressed due to the lower luminosity for sea quarks. The
current best bounds for the first generation quarks are given by
\cite{Allanach:1999ic, Barbier:2004ez,Chemtob:2004xr}.

\medskip

\begin{center}
\begin{tabular}{|c|ccc|}\hline
Coupling &$\;\lam'_{111}\;$& $\;\lam'_{211}\;$&$\;\lam'_{311}\;$ \\[1mm] \hline
best bound &$\;0.0005\;$&$\;0.059\;$ & 0.11\\ \hline
\end{tabular}
\end{center}

\medskip

The last two bounds scale with the relevant scalar fermion mass $(m_
{\tilde f}/100\,\mathrm{GeV})$. The first strict bound arises from
neutrinoless double beta decay searches and has a more complicated
$m_{\tilde f}$ dependence.

The outline of our paper is as follows. In
Sect.~\ref{sleptonproduction} we compute the NLO resonant slepton
production cross section at hadron colliders, including both virtual
and real diagrams. We also employ this calculation to determine the
running of the parameter $\lam'_{ijk}$. We then first compute the
partonic cross section and then numerically the hadronic cross
section. In Sect.~\ref{stm}, we investigate the transverse momentum
distribution of the produced sleptons at NLO. In particular we perform
a $p_T$ resummation of the gluon distribution and then compare with
the {\sc Herwig} and {\sc Susygen} Monte Carlo generators. In
Sect.~\ref{s-decay} we study the branching ratios of the various
slepton decays and determine the resulting like-sign dimuon event
rates at the LHC and the Tevatron. In Sect.~\ref{conclusion} we
summarize our results and conclude.


\section{Total Production Cross Section}
\label{sleptonproduction}
\cleqn
\subsection{LO Processes}
We first consider the single slepton production processes at leading
order, \figref{fig_feynmaenner}(a),
\begin{align}
\begin{split}
        d_k(p) + \bar{d}_j / \bar{u}_j (p')&\rightarrow
                \tilde{\nu}_i/ \tilde{\ell}_{Li} (q) \\
        \bar{d}_k(p') + d_j/u_j (p) &\rightarrow
                \tilde{\nu}^*_i / \tilde{\ell}^+_{Li} (q).
\label{fourprocesses}
\end{split}
\end{align}
The relevant parts of the Lagrangian for the production at hadronic
colliders follow from the superpotential \eqref{superpot}. Expressed
in terms of the (four-) component fields and using the projection
operators $P_{R/L} = \tfrac{1}{2}(1 \pm \gamma^5)$ we have
\begin{align}
\begin{split}
        \lan_{LQD}
        \supset & \lam_{ijk}^{\prime} \, 
                (\tilde{\nu}_i\,\bar{d}_k\, P_L\,  d_j
                - \tilde{\ell}_{Li}\, \bar{d}_k \, P_L\, u_j)\\
        &+ \lam_{ijk}^{\prime \ast}\, 
                (\tilde{\nu}^*_i \bar{d}_j\,P_R\, d_k 
                - \tilde{\ell}_{Li}^+ \, \bar{u}_j\, P_R\, d_k).
\label{eq_rpv_lan}
\end{split}
\end{align}
We obtain for the total partonic single slepton (charged slepton or
sneutrino) production cross section \cite{Dimopoulos:1988jw,
  Dimopoulos:1988fr}
\begin{align}
        \hat{\sigma}^{\rm LO} &=
                \frac{\abs{\lam^{\prime}_{ijk}}^2\,\pi}{12\hat s}
                        \,\,\delta(1-\tau) 
                \equiv \hat{\sigma}_0 \,\delta(1-\tau)
\label{eq_born1}
\end{align}
with $\tau= \tilde m^2/\hat s$, where $\tilde m$ is the slepton mass
and $\sqrt{\hat{s}}$ the partonic center-of-mass energy.

\subsection{NLO SUSY-QCD Corrections}
\label{Part_NLO}

\begin{figure}
        \subfigure[]{\scalebox{.8}{\epsffile{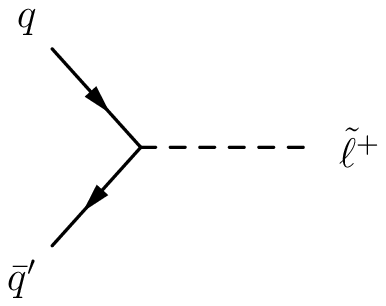}}}%
        \\[2ex]
        \subfigure[]{\scalebox{.8}{\epsffile{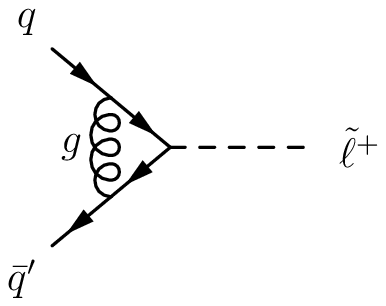}}}%
        \qquad%
        \subfigure[]{\scalebox{.8}{\epsffile{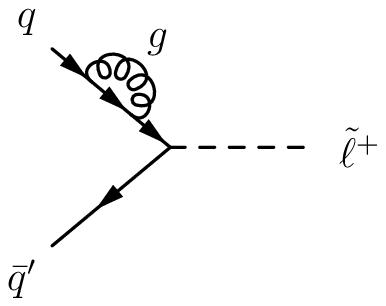}}}
        \\[2ex]
        \subfigure[]{\scalebox{.8}{\epsffile{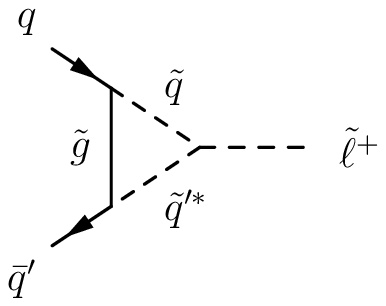}}}%
        \qquad%
        \subfigure[]{\scalebox{.8}{\epsffile{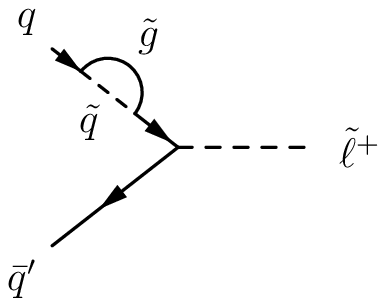}}}%
        \\[2ex]%
        \subfigure[]{\scalebox{.8}{\epsffile{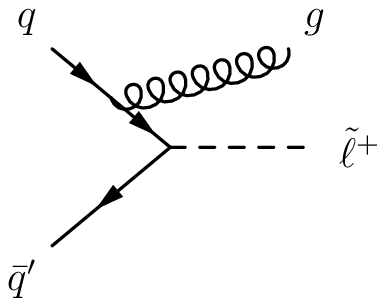}}}%
        \qquad%
        \subfigure[]{\scalebox{.8}{\epsffile{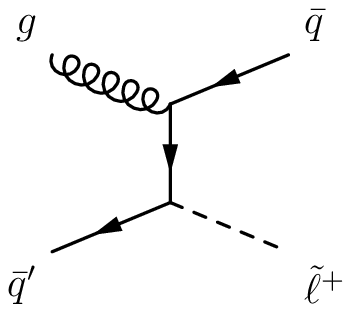}}}
\caption{Generic Feynman diagrams for resonant slepton production, 
  (a) LO; (b) QCD vertex corrections; (c) QCD self energy; (d)
  SUSY-QCD vertex correction; (e) SUSY-QCD self energy; (f) QCD real
  gluon radiation; (g) QCD Compton process.}
        \label{fig_feynmaenner}
\end{figure}

The cross section at ${\cal O}(\alpha_s)$ comprises radiative
corrections to the quark-antiquark process including virtual gluon,
gluino, quark and squark exchange,
\figref{fig_feynmaenner}(b),\,(c),\,(d),\,(e), real gluon radiation,
\figref{fig_feynmaenner}(f), and Compton-like gluon initiated
subprocesses, \figref{fig_feynmaenner}(g). In the following, we shall
concentrate on positively charged slepton production.  The production
of sneutrinos can be treated analogously. Throughout this section,
generation indices are suppressed.

We use dimensional regularization to isolate ultraviolet, infrared and
collinear divergences, and adopt the $\overline{\rm MS}$ scheme for
the renormalization of the coupling $\lambda^{\prime}$ and the
factorization of the collinear divergences through a redefinition of
the parton distribution functions.

The calculation of the NLO QCD corrections from virtual gluon exchange
and real gluon emission is straightforward. For the partonic
$q\bar{q}'$ cross section at NLO we find
\begin{widetext}
\begin{align}
\begin{split}           
  \hat{\sigma}^{\rm NLO, QCD}_{q\bar{q}'}=& \, 
        \hat{\sigma}_0\,\delta(1-\tau)\,\bigg\lbrace 
        1 + \frac{\alpha_s}{\pi} \,C_F\, \bigg(
          \frac{3}{2}\ln\frac{\mu_R^2}{\tilde{m}^2}
        - \frac{3}{2}\ln\frac{\mu_F^2}{\tilde{m}^2}
        + \frac{\pi^2}{3} -1 \bigg) \bigg\rbrace 
\\
  & +\, \hat{\sigma}_0\, 
        \frac{\alpha_s}{\pi}\,C_F\,\bigg\lbrace
        \frac{1+\tau^2}{[1-\tau]_+}\,
        \ln\frac{\tilde{m}^2}{\tau\,\mu_F^2}
        + 2(1+\tau^2)\,\bigg[ 
                \frac{\ln(1-\tau)}{1-\tau}\bigg]_+
                +(1-\tau) \bigg\rbrace,
\end{split}
\end{align}\end{widetext}
and for the quark-gluon scattering contribution
\begin{align}
\begin{split}
        \hat{\sigma}^{\rm NLO}_{qg}=& 
        \hat{\sigma}^{\rm NLO}_{\bar{q}'g}= 
                \hat{\sigma}_0 \, \frac{\alpha_s}{4\pi} \,
                \bigg\lbrace \frac{1}{2}\,(1-\tau)(7\tau-3)
\\
&               + \left[\tau^2 +(1-\tau)^2 \right]\, 
                \left[\ln\frac{\tilde{m}^2}{\mu_F^2} 
                      + \ln\frac{(1-\tau)^2}{\tau} \right]
                \bigg\rbrace ,
\end{split}
\end{align}
where $C_F=4/3$ and $\tau = \tilde m^2/\hat s$. 
The couplings $\lambda^{\prime}(\mu_R)$ and $\alpha_s(\mu_R)$ are
evaluated at the renormalization scale $\mu_R$, and $\mu_F$ denotes
the factorization scale. Our results agree with those presented in
\cite{Choudhury:2002au,Yang:2005ts}.

For a complete cross section prediction at ${\cal O}(\alpha_s)$ one
has to include SUSY-QCD corrections through virtual gluino and squark
exchange. These corrections have not been considered in previous
calculations. The SUSY-QCD corrections at NLO arise from the vertex
correction, \figref{fig_feynmaenner}(d), and the quark self energy,
\figref{fig_feynmaenner}(e), and are infrared-finite.

The interaction of quarks $q$, (right-) left-handed squarks
$\tilde{q}_L$ ($\tilde{q}_R$) and gluinos $\tilde{g}$ is described by
\cite{Haber:1984rc,Rosiek:1995kg}
\begin{align}
\lan_{q \tilde q \tilde g} = - & \sqrt{2} g_s\,t_{bd}^a\sum_{f=u,d}  
        \Big( \bar{\tilde{g}}_a \,P_L \,q_f^{d}\,\tilde{q}_{Lf}^{* b} 
                + \bar{q}_f^{b}\,P_R\, \tilde{g}_a\,\tilde{q}_{Lf}^{d}
\nonumber\\
        &       - \bar{\tilde{g}}_a\,P_R\,q_f^{d}\,\tilde{q}_{Rf}^{* b}
                - \bar{q}_f^{b}\,P_L\,\tilde{g}_a\, \tilde{q}_{Rf}^{d}
        \Big) .
\label{eq_qqg_int}
\end{align}
Here, $f$ denotes the quark flavor, $g_s$ the QCD coupling, $b,\,d$
($a$) are the indices of the fundamental (adjoint) representation of
$SU(3)_C$, and $t_{bd}^a$ are the generators of the group.  Squark
mixing between left- and right-handed squarks is neglected since only
light quarks in the initial state contribute to the hadron collider
cross section.  Moreover, mixing effects between squark generations
are excluded to avoid FCNC processes.

\subsubsection{Vertex correction}

For the SUSY-QCD vertex corrections, also the interaction vertices of
two squarks and a (left-handed) slepton are needed.  They can be
extracted from the $\mathrm{P}_6$-violating parts of the general soft
SUSY-breaking Lagrangian \cite{Barbier:2004ez}
\begin{align}
\begin{split}
\lan_{\tilde\ell \tilde q \tilde q}^{soft} \supset 
        -&\lamp_{ijk} A\,\left(
                \tilde{\nu}_i \,\tilde{d}_{Lj}\,\tilde{d}_{Rk}^* 
                - \tilde{\ell}_{Li}\,\tilde{u}_{Lj}\,\tilde{d}_{Rk}^*
                \right) 
\\
        &-\lam^{\prime \ast}_{ijk} A^{\ast}\, \left( 
                \tilde{\nu}_i^*\,\tilde{d}_{Lj}^*\,\tilde{d}_{Rk} 
        -\tilde{\ell}_{Li}^+\,\tilde{u}_{Lj}^*\,\tilde{d}_{Rk}
                \right).
\label{eq_softlagrangian}
\end{split} 
\end{align}
We have assumed here that the soft breaking terms have a universal
dimensionful parameter $A$ and are proportional to the dimensionless
coupling constant $\lamp_{ijk}$ of the superpotential
\cite{Soni:1983rm}.

Due to angular momentum conservation only those diagrams contribute
where the incoming quark and antiquark have the same helicity.
Therefore, for a specific slepton production process, only a single
vertex correction diagram contributes at NLO which is described by
\eqref{eq_softlagrangian}.

Using standard notation~\cite{Denner:1991kt} for scalar loop
integrals, the SUSY vertex contribution to the NLO cross section can
be written as
\begin{align}
\begin{split}
        & \hat{\sigma}^{\rm NLO}_{\rm SUSY,\,vc} = 
\\[1ex]
        & -\hat{\sigma}^{\rm LO}\,
                \frac{\alpha_s}{\pi}\, C_F\, A \,m_{\tilde g}\,  
                C_0\left(-p , p^{\prime},m_{\tilde g},
                m_{\tilde{q}_L},m_{\tilde{q}_R} \right)\,,
\label{eq_susyvertexcorr}
\end{split}
\end{align} 
where $p,p^{\prime}$ are the momenta of the incoming quark and
antiquark, $m_{\tilde g}$ is the gluino mass, and $m_{\tilde{q}_L},
m_{\tilde{q}_R}$ are the masses of the left- and right-handed squark
and antisquark running in the loop.

\subsubsection{Quark Self Energy}

SUSY-QCD self energy contributions arise from diagrams of the form
\figref{fig_feynmaenner}(e).  Since the SUSY partners of left- and
right-handed quarks can have different masses, the SUSY self energy
corrections to the left- and the right-handed quark field differ in
general.

The field strength renormalization constant $\tilde{Z}$ is
\begin{align}
        \tilde{Z}^{-1} &
           = 1- \Big(\tilde{\Sigma}_V - \tilde{\Sigma}_V\big\lvert_{UV}\Big)\,,
\end{align}
where the vector part of the self energy $\tilde{\Sigma}_V$ can be written
in terms of the coefficient function~$B_1$~\cite{Denner:1991kt} as
\begin{align}
        \tilde{\Sigma}_V &=  - \frac{\alpha_s}{2\pi}\, C_F\, 
                B_1(-p,m_{\tilde g},m_{\tilde{q}_{L/R}}).
\end{align}
The UV pole of the integral is, in the
$\overline{\rm MS}$ scheme,
\begin{align}
        B_1\lvert_{UV}&= - (4\pi)^{\eps}\,\Gamma(1+\eps) \, \frac{1}{2\eps},
\end{align}
where the number of space-time dimensions is $N=4-2\eps$.

The SUSY-QCD self energy contributions to the NLO production cross
section follow directly,
\begin{align}
\begin{split}
        &\hat{\sigma}^{\rm NLO}_{\rm SUSY,\,se} =  \hat{\sigma}^{\rm LO}\,
                \frac{\alpha_s}{2\pi}\, C_F\, 
\\              
&\times \Big( B_1(-p, m_{\tilde g},m_{\tilde{q}_L})
        + B_1(-p, m_{\tilde g},m_{\tilde{q}_R}) 
        - 2 B_1\lvert_{UV} \Big).
\label{eq_susyselfenergy}
\end{split}
\end{align}
This completes the computation of resonant slepton production at NLO.

\subsection{The running coupling $\lam^{\prime}$}

In this section, we briefly discuss the ${\cal O}(\alpha_s)$ running
of the $LQ\bar{D}$ coupling $\lambda'$ including both QCD and SUSY-QCD
corrections.

The bare $LQ\bar{D}$ coupling $\lam^{\prime 0}$ is related to the
renormalized coupling via
\begin{align}
        \lam^{\prime 0} = Z \lam^{\prime} \mu^{\eps},
\end{align}     
where $Z$ is the renormalization constant for the coupling $\lamp$ and
the scale $\mu$ is introduced to keep the renormalized coupling
dimensionless.  Note that the only scale dependence of $Z$ is carried
by $\alpha_s$ and, therefore,
\begin{align}
        \frac{dZ}{d\mu} &=
                \frac{dZ}{d\alpha_s}\,\frac{d\alpha_s}{d\mu} 
                = -\frac{\beta(\alpha_s)}{\mu} \, \frac{dZ}{d\alpha_s},
\end{align}
where the $\beta$-function is defined as $\beta(\alpha_s) = -\mu \,
d\alpha_s/d\mu$ with $\beta(\alpha_s)=2 \eps \alpha_s +
\mathcal{O}(\alpha_s^2)$.
We thus obtain for the $\beta$-function of the $LQ\bar{D}$ coupling
$\lamp$
\begin{align}
\begin{split}
        \beta(\lamp) \equiv& - \mu\, \frac{d \lamp}{d \mu} \\
        =& - \lamp \beta(\alpha_s) \frac{1}{Z} 
                        \frac{dZ}{d\alpha_s} + \lamp \eps.
\label{eq_beta_lam}
\end{split}
\end{align}

Considering only QCD corrections, the renorma\-li\-zation constant $Z$
depends on the (identical) renorma\-li\-zation constants for the
left-/right-handed quark field $Z_{L/R}$ and on the vertex
renormalization $Z_{\lamp}$. They are
\begin{align}
         Z_{\lamp} &= 1 -\, \frac{\alpha_s}{\pi} \, C_F \,(4\pi)^{\eps}\,
                \Gamma(1+\eps)\, \frac{1}{\eps}, \\
        Z_{L/R} &= 1 -\, \frac{\alpha_s}{4 \pi} \, C_F \,(4\pi)^{\eps}\,
                \Gamma(1+\eps)\, \frac{1}{\eps}.
\end{align}
Taking also the SUSY-QCD corrections into account, SUSY self energy
contributions $\tilde{Z}_{L/R}$ alter the renormalization constant
$Z$,
\begin{align}
        \tilde{Z}_{L/R} &= 1 - \frac{ \alpha_s}{4\pi}\,C_F\, 
                (4\pi)^{\eps}\, \Gamma(1+\eps)\, \frac{1}{\eps}.
\end{align}
Explicitely, including only QCD corrections,
\begin{align}
\begin{split}
Z       = \frac{Z_{\lamp}}{Z_L} 
        = 1 - \frac{3 \alpha_s}{4\pi}C_F\, 
                (4\pi)^{\eps}\,\Gamma(1+\eps)\, \frac{1}{\eps},
\end{split}
\end{align}
and including QCD and SUSY-QCD corrections,
\begin{align}
Z       =  \frac{Z_{\lamp}}{Z_L  \tilde{Z}_L}
        = 1 -\, \frac{\alpha_s}{2\pi}\, C_F\, 
                (4\pi)^{\eps}\,\Gamma(1+\eps)\, \frac{1}{\eps}.
\end{align}
For the $\beta$-function \eqref{eq_beta_lam} one obtains 
\begin{align}
        \beta(\lamp) &= 
                \big( \beta_{\lamp}^{\rm QCD} + \beta_{\lamp}^{\rm SUSY} 
                \big)\,\lamp \frac{\alpha_s}{2\pi} +
                \mathcal{O}(\lamp \alpha_s^2), 
\label{eq_beta_coeff}
\end{align}
with the coefficients $\beta_{\lamp}^{\rm QCD}, \beta_{\lamp}^{\rm SUSY}$ 
related to QCD and SUSY-QCD effects, respectively, 
\begin{align}   
        \beta_{\lamp}^{\rm QCD} = 3 C_F, \qquad \quad
        \beta_{\lamp}^{\rm SUSY} = - C_F,
\end{align} such that in total $\beta_{\lamp}^{\rm tot} =  
\beta_{\lamp}^{\rm QCD} + \beta_{\lamp}^{\rm SUSY} = 2 C_F$.

By integrating \eqref{eq_beta_lam} and inserting the definition
\eqref{eq_beta_coeff},  the NLO running of the $LQ\bar{D}$ coupling
is obtained as 
\begin{align}
  \lamp(\mu) &= \frac{\lamp(\mu_0)}{ 1 + \frac{\displaystyle
      \alpha_s}{\displaystyle 4\pi}\, \beta_{\lamp}^{\rm tot}\,
    \ln\frac{\displaystyle \mu^2}{\displaystyle \mu_0^2}},
\label{eq_lamrun}
\end{align}
in agreement with the results presented in
\cite{Martin:1993zk,Allanach:2003eb}; see however \footnote{We note
  that \cite{Alves:2002tj} reports a vanishing $\beta_{\lamp}^{\rm
    tot}$, in disagreement with our result and the results obtained in
  \cite{Martin:1993zk,Allanach:2003eb}.}.


\subsection{Decoupling of heavy SUSY particles}

In the limit of heavy SUSY particles $m_{\tilde q_L}, m_{\tilde q_R},
m_{\tilde g} \equiv \tilde M$ and $\tilde M \rightarrow \infty$, the
NLO SUSY-QCD vertex corrections vanish
\begin{align}
\hat{\sigma}_{\rm SUSY,\,vc}^{\rm NLO} &= -\hat{\sigma}_{\rm LO}\, 
\frac{\alpha_s}{\pi}\,
         C_F\, A\, \tilde{M}\, C_0(-p,p',\tilde{M},\tilde{M},\tilde{M}) 
        \nonumber\\
        &\rightarrow 0, 
\end{align}
while the self energy contributions are logarithmically enhanced
\begin{align}
 \hat{\sigma}_{\rm SUSY,\,se}^{\rm NLO} &= \hat{\sigma}_{\rm LO} \, 
\frac{\alpha_s}{2\pi}\,C_F\, 
         \big[ B_1(-p,\tilde{M}, \tilde{M}) + B_1(-p, \tilde{M},\tilde{M} ) 
\big] 
        \nonumber \\
        &\rightarrow - \hat{\sigma}_{\rm LO} \, \frac{\alpha_s}{2\pi}\,C_F\, 
\ln\frac{\mu^2}{\tilde{M}^2}.
\label{eq_susy_se}
\end{align}
The large logarithms can be absorbed by a redefinition of the
coupling $\lamp$ according to
\begin{align}
\begin{split}
  \lamp &= \lamp(\mu_0)\, \Big[ 1 - \frac{\alpha_s}{4\pi}\,
  \beta_{\lamp}^{\rm tot} \ln\frac{\mu^2}{\mu_0^2} \Big]\\
  &\downarrow \\
  \lamp + \Delta\lamp &= \lamp(\mu_0)\, \Big[1 -
  \frac{\alpha_s}{4\pi}\, \beta_{\lamp}^{\rm tot} \,
  \ln\frac{\mu^2}{\mu_0^2} - \frac{\alpha_s}{4\pi}\, C_F
  \ln\frac{\mu^2}{\tilde{M}^2} \Big].
\label{eq_dellamdef}
\end{split}
\end{align}
Then, the running of $\lamp$ is determined only by the QCD corrections
(light particles)
\begin{align}
\begin{split}
        \frac{ \del (\lamp + \Delta \lamp)}{\del \ln \mu^2} &= 
-\lamp(\mu_0)\, \frac{\alpha_s}{4\pi} \, 
                \big[ \beta_{\lamp}^{\rm tot} +C_F\big] \\
        &= -\lamp(\mu_0)\, \frac{\alpha_s}{4\pi}\, \beta_{\lamp}^{\rm QCD}.
\end{split}
\end{align}
Expanding $\lamp = (\lamp + \Delta\lamp) - \Delta\lamp$ in
$\hat{\sigma}_{\rm LO}$ in \eqref{eq_susy_se} cancels the term
$\propto \ln(\mu^2/\tilde{M}^2)$ so that in the limit $\tilde{M}
\rightarrow \infty$ the SUSY-QCD corrections vanish.  Note that we
will not adopt the decoupling scheme described above in the subsequent
numerical analysis as we do not consider scenarios where squark and
gluino masses are much larger than the slepton mass.


\subsection{Total Hadronic Cross Sections}

In this section, we will present numerical results for the total
hadronic cross section for resonant charged slepton production at the
Tevatron and at the LHC. We assume only a single non-zero coupling
$\lamp_{i11}=\lamp_{iud}$ and set this coupling $\lamp= 0.01$ at all
scales. The cross sections are directly proportional to $\lvert\lamp
\rvert^2$ such that results for other values of the coupling are 
easily obtained by rescaling. For the calculation of the $PP$ and
$P\bar P$ cross sections we have adopted the CTEQ6L1 and
CTEQ6M~\cite{Pumplin:2005rh} parton distribution functions at LO and
NLO, corresponding to $\Lam _5^{\mathrm{LO}}=165$~MeV and
$\Lam_5^{\overline{\mathrm{MS}}}=226\, $MeV at the one- and two-loop
level of the strong coupling $\alpha_s (\mu_R)$, respectively. The
renormali\-zation and factorization scales have been identified with the
slepton mass, $\mu_R = \mu_F = \tilde m$.

\begin{figure} 
\psfrag{tildem}[c][c]{\LARGE $\tilde m \quad $[TeV] }
\psfrag{Lhc}[l][lb]{\LARGE $\sqrt{S}=14$\,TeV ($PP$)}
\psfrag{Tev}[lt][lb]{\LARGE $\sqrt{S}=1.96$\,TeV ($P\bar{P})$}
\psfrag{sigma}[c][ct]{\LARGE $\sigma \quad $[pb]}
\psfrag{LO }[l][l]{\LARGE LO}
\psfrag{NLOQCD}[l][l]{\LARGE QCD }
\psfrag{NLOQCDSUSY}[l][l]{\LARGE QCD + SUSY (A=1\,TeV, $m_{\tilde q, 
\tilde g}=300$\,GeV)}
\psfrag{CTEQ6}{\LARGE CTEQ 6}
\psfrag{0}[r][r]{\LARGE 0}
\psfrag{0.5}[r][r]{\LARGE 0.5}
\psfrag{1}[r][r]{\LARGE 1}
\psfrag{1.5}[r][r]{\LARGE 1.5}
\psfrag{2}[r][r]{\LARGE 2}
\psfrag{0.0001}[r][r]{\LARGE $10^{-4}$}
\psfrag{0.001}[r][r]{\LARGE $10^{-3}$}
\psfrag{0.01}[r][r]{\LARGE $10^{-2}$}
\psfrag{0.1}[r][r]{\LARGE $10^{-1}$}
\psfrag{10}[r][r]{\LARGE 10 }
        \scalebox{.445}{\epsffile{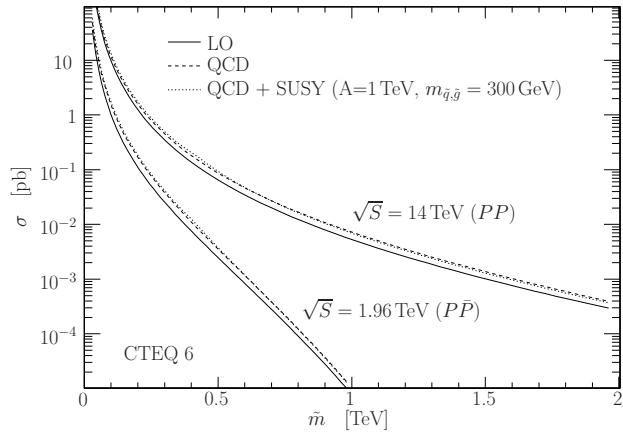}}
        \caption{Single slepton production cross section at
          the Tevatron (lower curves) and at the LHC (upper curves).
          The $LQ\bar{D}$-coupling is set to $\lamp_{i11} = 0.01$.
          The CTEQ6 LO/NLO PDFs have been used, and renormalization
          and factorization scales have been identified with the
          slepton mass $\tilde{m}$.}
\label{fig_crosssection}
\end{figure}

\begin{figure*}
\psfrag{tildem}[c][c]{$\tilde m \quad $[TeV] }
\psfrag{Lhc}[c][c]{$PP \rightarrow \tilde \ell_i^+ + X$}
\psfrag{Lhc2}[c][c]{\;\;\;$\sqrt{S}=14$\,TeV}
\psfrag{K-Factor}{$K$-factor}
\psfrag{QCD }[l][l]{QCD}
\psfrag{QCDSUSY0 }[l][l]{QCD+SUSY (A = 0 TeV)}
\psfrag{QCDSUSY1 }[l][l]{QCD+SUSY (A = 1 TeV)}
\psfrag{QCDSUSY-1 }[l][l]{QCD+SUSY (A = -1 TeV)}
\psfrag{massen1}[r][r]{$m_{\tilde q, \tilde g} = 300$\,GeV}
\psfrag{massen2}[r][r]{$m_{\tilde q, \tilde g} = 600$\,GeV}
\psfrag{massen3}[r][r]{$m_{\tilde q, \tilde g} = 1$\,TeV}
\psfrag{0.5}{0.5}
\psfrag{1}{1}
\psfrag{1.5}{1.5}
\psfrag{2}{2}
\psfrag{1}{1 }
\psfrag{1.1}{1.1 }
\psfrag{1.2}{1.2 }
\psfrag{1.3}{1.3 }
\psfrag{1.4}{1.4 }
\psfrag{1.5}{1.5 }
        \scalebox{.82}{\epsffile{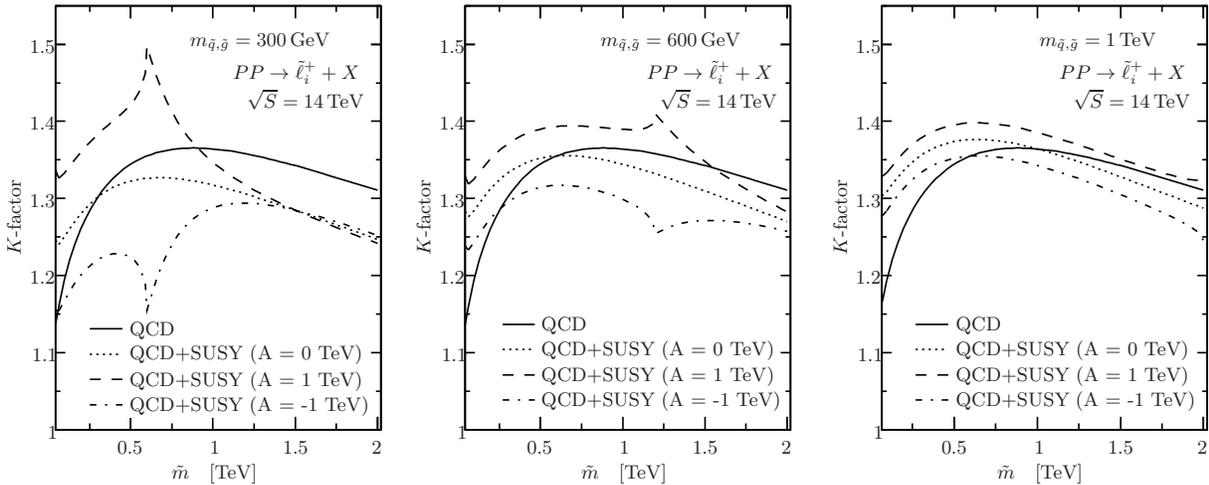}}
        \caption{SUSY-QCD $K$-factor for resonant slepton production
        via $\lambda'_{i11}=0.01$ at the LHC, $\sqrt{S}=14$\,TeV, for
        three sets of squark and gluino masses ($m_{\tilde q_L} =
        m_{\tilde q_R} = m_{\tilde g}$). Different values for the SUSY
        breaking parameter $A$ are chosen as indicated. For comparison
        we give also the QCD $K$-factor (solid lines).  The CTEQ6
        LO/NLO PDFs have been used, and renormalization and
        factorization scales have been identified with the slepton
        mass $\tilde{m}$.}
\label{fig_Kfact_LHC}
\end{figure*}

\begin{figure*} 
\psfrag{tildem}[c][c]{$\tilde m \quad $[TeV] }
\psfrag{Tev}[c][l]{$P\bar{P} \rightarrow \tilde \ell_i^+ + X$}
\psfrag{Lhc2}[c][c]{$\sqrt{S}=1.96$\,TeV}
\psfrag{K-Factor}{$K$-factor}
\psfrag{QCD }[l][l]{QCD}
\psfrag{QCDSUSY0 }[l][l]{QCD+SUSY (A = 0 TeV)}
\psfrag{QCDSUSY1 }[l][l]{QCD+SUSY (A = 1 TeV)}
\psfrag{QCDSUSY-1 }[l][l]{QCD+SUSY (A = -1 TeV)}
\psfrag{massen1}[r][r]{$m_{\tilde q, \tilde g} = 300$\,GeV}
\psfrag{massen2}[r][r]{$m_{\tilde q, \tilde g} = 600$\,GeV}
\psfrag{massen3}[r][r]{$m_{\tilde q, \tilde g} = 1$\,TeV}
\psfrag{0}{0}
\psfrag{0.2}{0.2}
\psfrag{0.4}{0.4}
\psfrag{0.6}{0.6}
\psfrag{0.8}{0.8}
\psfrag{1}{1 }
\psfrag{1.2}{1.2 }
\psfrag{1.4}{1.4 }
\psfrag{1.6}{1.6 }
\psfrag{1.8}{1.8 }
        \scalebox{.82}{\epsffile{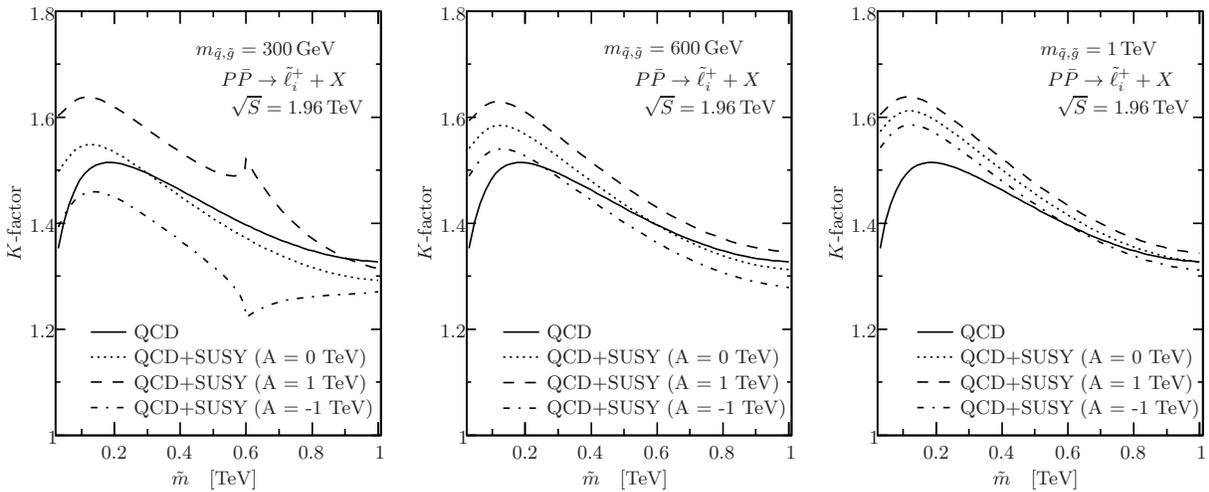}}
        \caption{The same as \figref{fig_Kfact_LHC} but for the Tevatron.}
\label{fig_Kfact_Tev}
\end{figure*}

The hadronic production cross section for the process $u \bar{d}
\rightarrow \tilde{\ell}_i^+$ at the Tevatron and at the LHC is shown
as a function of the slepton mass in \figref{fig_crosssection}. We
compare the LO result with the NLO results with only QCD and both QCD
and SUSY-QCD corrections included.  The NLO cross section for
producing sleptons with a mass $\tilde{m}= 100$\,GeV is approximately
1~pb at the Tevatron and 10~pb at the LHC, decreasing rapidly with
increasing slepton mass. The slepton discovery reach at the Tevatron
is thus limited by the comparably small cross section, with only one
expected event at $\tilde m=600$\,GeV (for $\lambda'_{i11}=0.01$) for
an integrated luminosity of $\int \! dt\, {\cal L} = 1$\,fb$^{-1}$.
The cross section is more favourable at the LHC, where a 600\,GeV
slepton would be produced with $\sigma \approx 50$\,fb, leading to
500~events per year even for the low luminosity phase with ${\cal L} =
10$\,fb$^{-1}/$year.

To demonstrate the impact of the SUSY-QCD corrections more clearly, we
consider the $K$-factor $K=\sigma_{\rm NLO}/\sigma_{\rm LO}$ defined
as the ratio of NLO to LO cross section, with all quantities
calculated consistently in lowest and next-to-leading order.  In
\figref{fig_Kfact_LHC} (\ref{fig_Kfact_Tev}) we display the QCD as
well as the SUSY-QCD $K$-factor at the LHC (Tevatron) for three
different sets of squark and gluino masses and for three different
values of the SUSY breaking parameter~$A$.

The QCD $K$-factor shows an enhancement of the LO cross section of up
to $35\%$ at the LHC and of up to $50\%$ at the Tevatron. For
scenarios where the trilinear SUSY breaking parameter $A$ vanishes,
the SUSY-QCD corrections are moderate. They enhance the cross section
prediction by up to $10\,\%$ for light sleptons and reduce it by
$5-10\,\%$ for heavier sleptons. For non-vanishing $A$ there are
additional contributions due to the SUSY-QCD vertex corrections which
can lead to a significant further enhancement or reduction of the
cross section by up to 15\% for $m_{{\tilde q},{\tilde g}}=300\,$ GeV.
For light squark masses, it can be as much as 35\%. One observes the
usual resonance peak when the slepton mass is twice the mass of the
squarks in the vertex loop.

We remark that the inclusion of the NLO-QCD corrections reduces the
factorization scale dependence of the theoretical prediction as
expected. At the Tevatron, we consider the slepton mass interval from
100 GeV to 1 TeV. A variation of the factorization scale between
$\mu_F = 2\tilde{m}$ and $\mu_F=\tilde{m} /2$ then changes the LO
cross section by up to $\pm$20\%. This uncertainty is reduced to less
than $\pm$10\% at NLO. At the LHC, we consider slepton masses between
100 GeV and 2 TeV and again vary the factorization scale between
$\mu_F = 2\tilde{m}$ and $\mu_F=\tilde{m} /2$. The LO uncertainty is
then up to $\pm$10\% which is reduced to less than $\pm$5\% in NLO.


\begin{table*}[tb]\begin{center}
    \scalebox{.95}{
\begin{tabular}{|c|c|c|c|c||c|c|c|c|}
\hline\rule[-3mm]{0mm}{8mm}
&\multicolumn{4}{c||}{CTEQ 6}&\multicolumn{4}{c|}{MRST 04 }\\[1ex]
\cline{2-5} \cline{6-9}
&&&\multicolumn{2}{c||}{SUSY-QCD} &&&\multicolumn{2}{c|}{SUSY-QCD}\\
&\rb{\; LO \;}& \rb{\; QCD \;} & $A = 0 $\,TeV& $A = 1$\,TeV%
&\rb{\; LO \;}& \rb{\; QCD \;}&$A =
0$\,TeV& $A = 1$\,TeV\\
\hline&&&&&&&&\\
$\sigma$\, [fb] &25.72  &38.43  &39.23  &40.22  &25.55  &37.73  &38.52
&39.50\\
Tevatron        &&&&&&&&\\
&&&&&&&&\\
$\sigma$\, [fb]&358.7   &467.0  &478.0  &491.6  &347.4  &471.2  &482.4
&496.2\\
LHC             &&&&&&&&\\[.5ex]
\hline
\end{tabular}}
\caption{\label{table:numericalresults} Numerical results for the 
  production cross section for a slepton of mass $\tilde m=300$\,GeV
  at the Tevatron ($\sqrt{S} = 1.96$\,TeV) and the LHC ($\sqrt{S} =
  14$\,TeV). Slepton production via the $u\bar{d}$ initial state is
  considered, with $\lambda'_{i11} = 0.01$.  All scales are fixed to
  the central scale $\mu= \tilde m$.  Squark and gluino masses are
  chosen as $m_{\tilde q_L} = m_{\tilde q_R} = m_{\tilde g}=
  600\,$GeV. Results are presented for the
  CTEQ6L1/CTEQ6M~\cite{Pumplin:2005rh} and the
  MRST2002LO/MRST2004NLO~\cite{Martin:2002dr,Martin:2004ir} PDF
  parametrizations.}
\end{center}       
\end{table*}

We conclude this section by tabulating some selected cross section
predictions for slepton production with $\tilde{m} = 300$~GeV at the
Tevatron or the LHC via the $u\bar{d}$ initial state in
Table\,\ref{table:numericalresults}. We show results for the default
CTEQ6M parton density functions (PDFs) and for the
MRST~\cite{Martin:2002dr, Martin:2004ir} parametrization.  We find
that the difference between the two PDF parametrizations is less than
approximately $5\%$ for the NLO cross section predictions.


\section{Slepton Transverse Momentum 
Distribution \label{stm}}

\cleqn

In this section we compute the transverse momentum ($p_T$)
distribution of the produced sleptons. At Born level, momentum
conservation enforces the produced sleptons to have zero $p_T$. At $
\mathcal{O}(\alpha_s)$ the sleptons may have non-zero $p_T$ due to an
additional parton in the final state.  The differential hadronic cross
section in $p_T$ at $\mathcal{O}(\alpha_s)$ is given in Appendix
\ref{pertpT}.  However, for $p_T \ll \tilde m$, with $\tilde m$ the
slepton mass, the emitted parton becomes soft and/or collinear to one
of the incoming partons leading to large logarithms in the
differential cross section. The perturbation series in $\alpha _s$
breaks down and has to be replaced by a series in $\alpha_s^n\log
^m(\tilde m^2/p_T^2)$ with $m,n\in \mathbb{N}$ and $m=0,\ldots, 2n-1$.
We shall employ the Collins, Soper, and Sterman (CSS) formalism to
resum the large logarithms \cite{Collins:1981uk,Collins:1981va,
  Collins:1984kg}.

The $p_T$~spectrum of the sleptons has also been studied in
\cite{Yang:2005ts}. We present the main formul{\ae} and extend the
studies of \cite{Yang:2005ts} by a comparison with Monte Carlo
simulations.

\subsection{Soft and Collinear Gluon Summation}
The summed differential cross section, valid in the region of soft
and/or collinear emitted partons, is given by \cite{Collins:1984kg}
\begin{align}
        \frac{d^2 \sigma^{\text{resum}}}{dp_T dy} &=
        \frac{\hat{\sigma}_0 p_T \tilde m^2}{S}
        \int_0^{\infty} db\,
        b\, J_0(bp_T) W(b) \, ,
\label{resum}
\end{align}
where $S$ is the hadronic c.m. energy, $J_0$ the zeroth order Bessel
function, and $\hat\sigma_0=\hat\sigma_0(\hat{s}=\tilde{m}^2)$ is the
leading order cross section defined in Eq.~(\ref{eq_born1}).  The
Sudakov-like form factor $W(b)$ is given by
\begin{align}
        W(b) =& \exp \bigg \lbrace\!\!-\!\!\int_{(b_0/b)^2}^{\tilde m^2} 
                \frac{dq^2}{q^2}\, \frac{\alpha_s(q^2)}{2\pi} 
                \Big[ K_1 \ln\frac{\tilde m^2}{q^2} + K_2 \Big] \bigg\rbrace
                \nonumber \\
             &\times \bigg[ f_{q A}\big( x_1^0\big) 
                f_{\bar{q}'B}\big(x_2^0\big)
                + \big( x_1^0 \leftrightarrow x_2^0 \big) \bigg] \,,
\label{resum1}
\end{align}
where $x_{1/2}^0 \equiv e^{\pm y}\sqrt{\tilde{m}^2/S}$ are the parton
momentum fractions in the limit $p_T/\tilde{m} \rightarrow 0$ and $y$
denotes the slepton rapidity. The $f_{qA},\,f_{\bar q ' B}$ are the
parton density functions for hadrons $A,\,B$, respectively. The
coefficients $K_1,\,K_2$ are in general functions of $\alpha_s$.
However, as we discuss below, we are here only interested in the
exponential at $\mathcal{O}(\alpha_ s)$, \textit{i.e.} $K_{1,2}$ are
constant.  The integration in Eq.~(\ref{resum1}) has to be performed
over the impact parameter $b$, the Fourier conjugate of $p_T$.  We
perform the integration analytically by using the LO running of
$\alpha_s$.  The PDF's are evaluated at a fixed factorization scale
$\mu_F$, where the most convenient choice \cite{Collins:1984kg} is
$\mu_F=b_0/b$ with $b_0=2e^{-\gamma_E}$ and $\gamma_E$ the
Euler-Mascheroni constant.

We calculate the coefficients $K_{1,2}$ by expanding (\ref{resum}) up
to $\mathcal{O}(\alpha_s)$. After some algebra one obtains
\cite{Han:1991sa}
\begin{widetext}
\begin{align}
        \frac{d^2\sigma^{\text{asym}}}{dp_T\, dy} =& \, 
                \frac{\hat{\sigma}_0 \,{\tilde m^2}}{S\, p_T}\,
                \frac{\alpha_s}{\pi}\,
        \bigg\lbrace \left[ K_1 \ln\frac{\tilde m^2}{p_T^2} +K_2\right]
                \,f_{q A}( x_1^0) \, f_{\bar{q}'B}(x_2^0)
\nonumber \\
        & +  \sum_{i=\bar{q}',g}
                 f_{q A}(x_1^0)\,\big(P_{\bar{q}'i}
                        \circ f_{i B}\big)(x_2^0)
        + \sum_{i=q,g}
                f_{\bar{q}' A}(x_1^0)\, \big( P_{qi}
                        \circ f_{i B} \big) (x_2^0) 
        \bigg\rbrace
                 + \big(x_1^0 \leftrightarrow x_2^0 \big)\, ,
\label{asym}
\end{align}
\end{widetext}
where $P_{\bar{q}'i},\,P_{qi}$ are the Altarelli-Parisi splitting
functions \cite{Altarelli:1977zs}, and the convolution is defined as
\begin{align}
        \Big(f \circ g\Big) (x) &\equiv \int_x^1\!\frac{dz}{z}\,
                 f(z)\,g\Big(\frac{x}{z}\Big) \, .
\end{align}
We can then compare the soft- and collinear divergences of this
asymptotic cross section with those of the perturbative result in the
limit $p_T\rightarrow0$.  We obtain $K_1=2C_F$ and $K_2=-3C_F$ with
$C_F =4/3$.

It was pointed out in \cite{Parisi:1979se} that $W(b)$ is ill-defined
when $b > 1/\Lambda_{QCD}$, because confinement sets in.  We factor
out the non-perturbative part by replacing $W(b)$ by
\cite{Collins:1984kg,Davies:1984sp}
\begin{align}
        W(b) & \rightarrow W(b_*)\, e^{-S_{np}(b)} \, , 
        \label{nonpert} \\
        b_* & \equiv \frac{b}{\sqrt{1+b^2/b^2_{\text{max}}}} \, ,
\end{align}
which smoothly cuts-off the region $b > b_{\text{max}}$, and our choice
of $b_{\text{max}}$ is given below. $S_{np}(b)$ parameterizes the
non-perturbative part and has to be determined by experiment.  We
follow the approach of \cite{Landry:2002ix}, where $e^{-S_{np}(b)}$ is
approximated by a Gaussian with adequate parameters fitted to
Drell-Yan data resulting in
\begin{align}
        S_{np}(b) &= b^2 \Big[ g_1 + g_2 \ln\frac{\tilde m}{2Q_0} + g_1g_3
                \ln(100 \,x_1^0\, x_2^0) \Big] \, .
\label{Gaussian_damping}
\end{align}
With $b_{\rm max} = 0.5\,$GeV$^{-1}$ and $Q_0=1.6\,$GeV,
\cite{Landry:2002ix} finds $g_1 = 0.21^{+0.01}_{-0.01}\,\text{GeV}^2$,
$g_2=0.68^{+0.01}_{-0.02}\,\text{GeV}^2$, $g_3=-0.60^{+0.05}_{-0.04}$.
Yang et al. \cite{Yang:2005ts} used the approach of \cite{Qiu:2000hf}
to factor out nonperturbative physics.

\subsection{Matching}
At low $p_T$, the summed cross section describes the spectrum
accurately compared to the perturbative cross section at
$\mathcal{O}(\alpha_s)$, \eqref{pert}.  All terms singular in $p_T$
are included.  For large $p_T$, also the non-singular terms must be
considered.  They correspond to the difference of the perturbative and
the asymptotic cross section at $\mathcal{O}(\alpha_s)$. For large
$p_T$ the summed (\ref{resum}) and asymptotic (\ref{asym}) cross
sections are negative, which is unphysical.  Therefore, we suppress
their contributions with an empirical matching function
\cite{Hinchliffe:1988ap,Kauffman:1991jt}:
\begin{align}
\frac{d^2\sigma^{\text{full}}}{dp_Tdy} = &
        \frac{1}{1+(p_T/p_T^{\text{match}})^4} \left[\frac{d^2\sigma^
  \text{resum}}
        {dp_Tdy} - \frac{d^2\sigma^{\text{asym}}}{dp_Tdy} \right] 
        \nonumber \\
        & + \frac{d^2\sigma^{\text{pert}}}{dp_Tdy} \, .
\label{fullpT}
\end{align} 
We choose $p_T^{\text{match}}=\tilde m/6$ at the Tevatron and
$p_T^{\text{match}}={\tilde m}/3$ at the LHC.  

\subsection{Results}

\begin{figure}[ht!] 
  \setlength{\unitlength}{1cm} \includegraphics[scale=0.6, bb = 22 60
  720 530, clip=true]{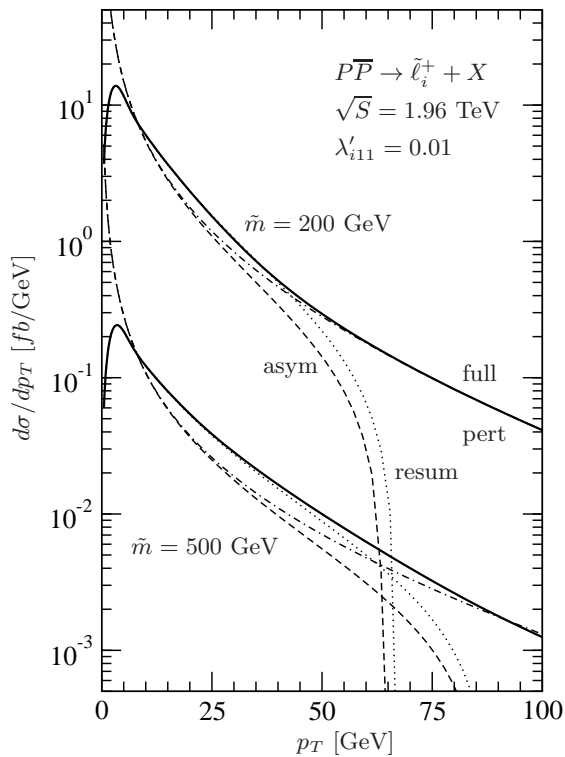}
  \put(-10.2,8.8){$P\overline{P} \rightarrow \tilde{\ell}^+_i + X$}
  \put(-10.2,8.3){$\sqrt{S}=1.96 \,\,\rm{TeV}$}
  \put(-10.2,7.8){$\lambda'_{i11}=0.01$}
  \put(-11.4,6.8){$\tilde{m}=200 \,\, \rm{GeV}$}
  \put(-12.9,2.5){$\tilde{m}=500 \,\, \rm{GeV}$}
  \put(-11.15,4.9){asym} \put(-9.4,3.5){resum} \put(-8.5,4.0){pert}
  \put(-8.5,4.8){full} \put(-10.7,-0.1){$p_T$ [GeV]}
  \put(-14.5,4.0){\rotatebox{90}{$d\sigma/dp_T$ [$fb$/GeV]}}
        \caption{\label{pTspectrum_Tev}
          Differential transverse momentum distribution of the slepton
          at the Tevatron for the $u\overline{d}$ initiated process
          via $\lambda'_{i11}$.  The two sets of curves correspond to
          slepton masses of $200$ and $500\,{\rm GeV}$. The different
          lines of each set denote the perturbative (dot dashed),
          asymptotic (dashed), summed (dotted) and full (solid)
          differential cross section.}
\end{figure}

\begin{figure}[ht!] 
  \setlength{\unitlength}{1cm} \includegraphics[scale=0.6, bb = 22 60
  720 530, clip=true]{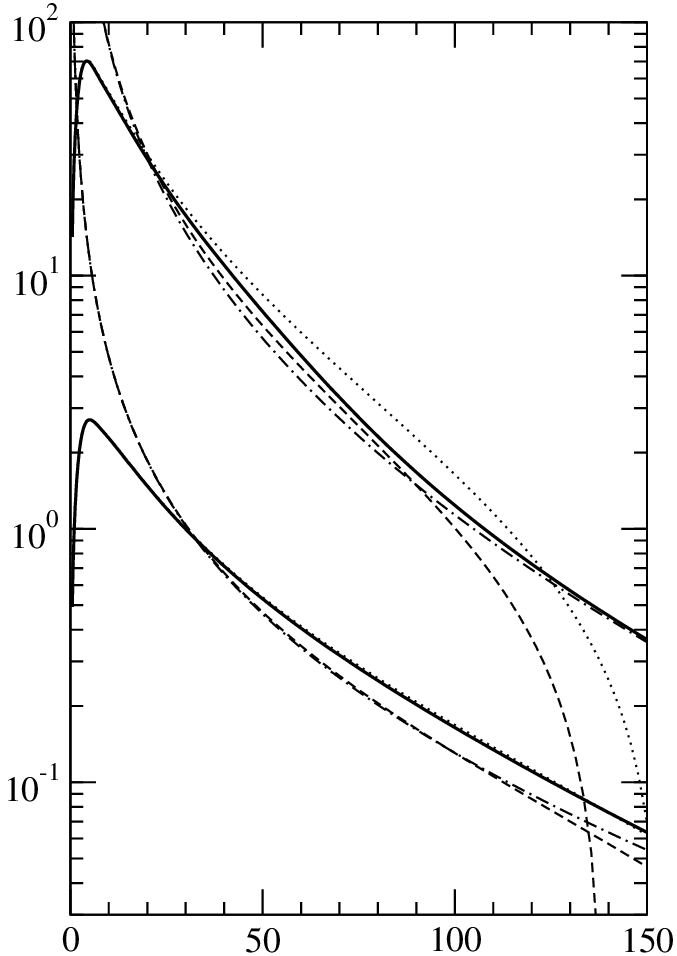}
  \put(-10.2,8.8){$P P \rightarrow \tilde{\ell}^+_i + X$}
  \put(-10.2,8.3){$\sqrt{S}=14 \,\,\text{TeV}$}
  \put(-10.2,7.8){$\lambda'_{i11}=0.01$}
  \put(-10.9,6.8){$\tilde{m}=200 \,\, \text{GeV}$}
  \put(-12.3,2.5){$\tilde{m}=500 \,\, \text{GeV}$}
  \put(-10.7,-0.1){$p_T$ [GeV]}
  \put(-14.5,4.0){\rotatebox{90}{$d\sigma/dp_T$ [$fb$/GeV]}}
        \caption{\label{pTspectrum_LHC}
          Same as \figref{pTspectrum_Tev} but for the LHC.}
\end{figure}

The full transverse momentum distribution for the case of a
(left-handed) slepton at the Tevatron (LHC) is shown in
\figref{pTspectrum_Tev} (\figref{pTspectrum_LHC}). The $y$-dependence
has been integrated out and the renormalization and factorization
scales of the asymptotic and perturbative cross section have been set
equal to $\mu = \sqrt{p_T^2+{\tilde m}^2}/2$.  Again we set
$\lambda'_{i11}= 0.01$ and use the CTEQ6M PDFs \cite{Pumplin:2005rh}.
We normalize the integral of the transverse momentum distribution to
the total hadronic cross section including QCD corrections. In the
low-$p_T$ region, the perturbative and the asymptotic cross section
cancel each other and the spectrum is described well by the summed
prediction. At high $p_T$, \textit{i.e.}  $p_T>50\,{\rm GeV}$
($100\,$GeV) at the Tevatron (LHC), the summed and asymptotic
contribution start to fall off and the distribution is to a good
approximation given by the perturbative cross section.  The summed
cross section peaks at $p_T = 3.1\,{\rm GeV}$ ($p_T = 3.4\,{\rm GeV}$)
at the Tevatron and at $p_T = 4.3\,{\rm GeV}$ ($p_T = 5.0\,{\rm GeV}$)
at the LHC for ${\tilde m}=200\,{\rm GeV}$ (${\tilde m}=500\,{\rm
GeV}$)

Yang et al.\ \cite{Yang:2005ts} derived the transverse momentum
distribution for sneutrinos with masses of $200\,$GeV, $400\,$GeV and
$600\,$GeV.  They did not use a matching procedure (\ref{fullpT}) and
did not normalize the integral of the transverse momentum distribution
to the total hadronic cross section including QCD corrections. To
facilitate the comparison with \cite{Yang:2005ts} we have also
calculated the $p_T$ distribution of the sneutrinos without a matching
function and without normalizing to the total NLO cross section. In
the phenomenologically relevant region, \textit{i.e.} where $p_T \lsim
50\,$GeV, there is good agreement with \cite{Yang:2005ts}. At high
$p_T$ we find discrepancies.  Our differential cross section at $p_T =
100\,$ GeV ($p_T = 200\,$GeV) is $75\%$ ($30\%$) larger (smaller)
compared to \cite{Yang:2005ts} for a sneutrino with a mass of
$200\,$GeV at the Tevatron (LHC). These discrepancies are not
particularly relevant phenomenologically, because the differential
cross section is between two and three orders of magnitude smaller at
$p_T = 100\,$ GeV ($p_T = 200\,$GeV) at the Tevatron (LHC) compared to
the peak region.

Within this computation, there are some sources of theoretical error.
First, the choice of the scales $\mu_F$ and $\mu_R$ and the empirical
nature of the matching procedure is ambiguous.  These uncertainties
mainly effect the high-$p_T$ region, where there are only few events.
Second, the parameters describing non-perturbative physics, \textit{
  cf.}  \eqref{Gaussian_damping}, are also not unique. Third, there
are uncertainties from the choice of the PDFs. In order to quantify
the latter, we have adopted the MRST2004NLO~\cite{Martin:2004ir} PDF
set for comparison. The maximal difference between the two $p_T$
distributions is at the peak.  It is less than $0.9\%$ ($4.6\%$) at
the Tevatron (LHC) for $\tilde{m}=200\,$GeV. Finally, there is of
course an error due to higher order corrections of $\mathcal{O}(
\alpha_s^2)$.

\subsection{Comparison with Monte Carlo Generators}

\begin{figure}[ht!] \centering
  \setlength{\unitlength}{1cm} \includegraphics[scale=0.6, bb = 22 210
  720 530, clip=true]{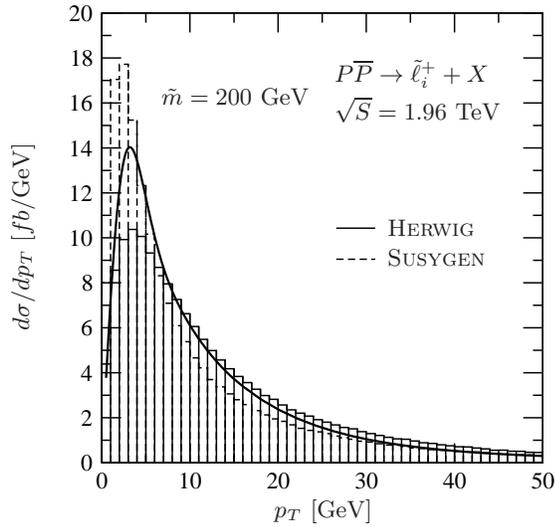}
  \put(-12.5,5.35){$\tilde{m} = 200 \,\, \rm{GeV}$}
  \put(-10.2,5.65){$P \overline{P} \rightarrow \tilde{\ell}^+_i + X$}
  \put(-10.2,5.15){$\sqrt{S}=1.96 \,\,\rm{TeV}$}
  \put(-9.5,3.6){{\sc Herwig}} \put(-9.5,3.2){{\sc Susygen}}
  \put(-14.5,2.35){\rotatebox{90}{$d\sigma/dp_T$ [$fb$/GeV]}}
  \put(-11.0,-0.15){$p_T$ [GeV]}
        \caption{\label{Comparison_with_MC_Tev}
          Transverse momentum distribution of the slepton at the
          Tevatron generated with {\sc Herwig} (solid histogram) and
          {\sc Susygen} (dashed histogram) compared with the
          CSS formalism (solid line).  The slepton mass is set to
          $200$~GeV, and the curves are normalized to the NLO QCD
          cross section.}
\end{figure}

\begin{figure}[ht!] \centering
\setlength{\unitlength}{1cm}
\includegraphics[scale=0.6, bb = 22 210 720 530, clip=true]{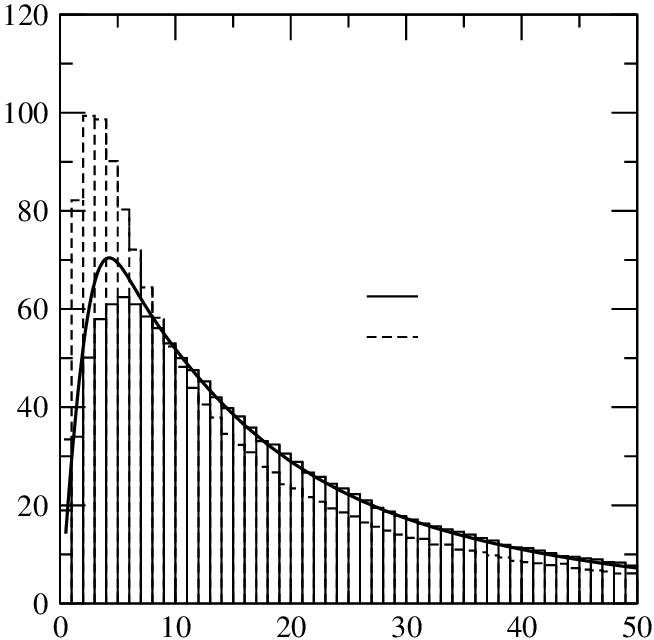}
        \put(-12.5,5.35){$\tilde{m} = 200 \,\, \rm{GeV}$}
        \put(-10.2,5.65){$P P \rightarrow \tilde{\ell}^+_i + X$}
        \put(-10.2,5.15){$\sqrt{S}=14 \,\,\rm{TeV}$}
        \put(-9.5,3.6){{\sc Herwig}}
        \put(-9.5,3.2){{\sc Susygen}}
                \put(-14.5,2.35){\rotatebox{90}{$d\sigma/dp_T$ [$fb$/GeV]}}
        \put(-11.0,-0.15){$p_T$ [GeV]}
        \caption{\label{Comparison_with_MC_LHC}
          Same as \figref{Comparison_with_MC_Tev} but for the LHC.}
\end{figure}

Next, we compare our analytic results for the slepton transverse
momentum distribution with the distributions predicted by two Monte
Carlo event generators: {\sc Herwig}~6.510~\cite{Marchesini:1991ch,
  Corcella:2000bw,Corcella:2002jc,Moretti:2002eu} and {\sc
  Susygen}~3~\cite{Ghodbane:1999va}, which uses the {\sc
  Pythia}~6.205~\cite{Sjostrand:2001yu,Mrenna:1996hu} parton shower.
The Monte Carlo programs generate the $p_T$ distribution from the LO
process through the parton shower, for which we use the default
settings.  In \figref{Comparison_with_MC_Tev}
(\figref{Comparison_with_MC_LHC}) we compare the transverse momentum
distribution as predicted by the analytic resummation with {\sc
  Herwig} and {\sc Susygen}. All curves are normalized to the NLO QCD
cross section. For both the Tevatron and the LHC, the spectrum
generated with {\sc Susygen} (through the {\sc Pythia} parton shower)
is significantly softer than that of {\sc Herwig}. The analytic
summation lies in between the two Monte Carlo programs. The
discrepancy between the {\sc Herwig} and {\sc Pythia} parton showers
is a known problem, which also occurs, for example, for resonant
single $Z$-boson production \cite{Peter}. Our analytic results might
help to improve future versions of Monte Carlo generators for resonant
scalar particle production.

\section{Slepton Decay}
\label{s-decay}

\cleqn

\begin{table*}[tb]
\begin{ruledtabular}
\begin{tabular}{c|r r r r r r r r r r r r}
Points & $\tilde{\mu}_L$ & $\tilde{\nu}_\mu$ & $\tilde{\tau}^1$ & 
$\tilde{\tau}^2$ & $\tilde{\nu}_\tau$ & $\tilde{\chi}_1^0$ & 
$\tilde{\chi}_2^0$ & $\tilde{\chi}_3^0$ & $\tilde{\chi}_4^0$ 
& $\tilde{\chi}_1^+$ & $\tilde{\chi}_2^+$ & $h^0$\\
\hline
\hline
1a  & 202 & 186 & 134 & 206 & 185 & 97 & 181 & - & - & 180 & - & - \\
\hline
1b  & 339 & 329 & 197 & 345 & 318 & 162 & 307 & - & - & 307 & - & 115 \\
\hline
2   & 1459 & 1456 & 1440 & 1453 & 1450 & 123 & 235 & 478 & 492 & 235 & 493 
& - \\
\hline
3   & 288 & 277 & 173 & 290 & 276 & 161 & - & - & - & - & - & 114 \\
\hline
4   & 449 & 441 & 260 & 415 & 388 & 121 & 226 & 401 & 416 & 226 & 418 & 113 \\
\hline
5           & 257 & 245 & 182 & 258 & 243 & 120 & 230 & - & - & 230 & - & -\\
\end{tabular}
\caption{\label{SUSYparam}
        Relevant masses of the mSUGRA spectrum for the six Snowmass 
        points, \textit{i.e.} those which are kinematically accessible in
        the slepton decays. All masses are in GeV, rounded to the nearest 
        whole number.}
\end{ruledtabular}
\end{table*}

In order to determine the final state particles observed in a collider
experiment as well as their event rates, we must discuss the possible
decays of the resonantly produced (left-handed) sleptons.  Since all
SUSY particles, even the LSP, are unstable in $\mathrm{P}_6$-violating
SUSY models, their decay signatures are crucial for experimental
analyses. Here we shall assume a neutralino LSP. Note that in $\mathrm
{P}_6$-violating SUSY models one can also have a scalar tau LSP which
dramatically alters the potential signatures~\cite{Allanach:2003eb}.

Apart from the inverted production process, the sleptons can decay
through gauge interactions.  Neglecting mixing between left- and
right-handed sleptons the dominant decays are, cf. \eqref{slep}:
\begin{align}
\begin{split}
        d_j {\bar d}_k\ra{\tilde\nu^*}_i&\ra
        \left\{ 
                \begin{array}{cc}
        d_j {\bar d}_k,          & \\
        {\bar \nu}_i\tilde\chi^0_m,    & \\
        \ell^+_i\tilde\chi^-_n, &
                \end{array}
        \right.\,,
\\
        u_j{\bar d}_k \ra{\tilde\ell}_i^+&\ra
        \left\{ 
                \begin{array}{ccc}
        u_j{\bar d}_k,          & \\
        \ell^+_i\tilde\chi^0_m, & \\
        {\bar \nu}_i\tilde\chi^+_n,    & 
                \end{array}
        \right.\,.
\label{slep2}
\end{split}
\end{align}

In order to quantitatively discuss the possible signal rates, we
consider the parameters of the six mSUGRA SPS points \cite{
Allanach:2002nj}. We calculate the spectra using the {\sc
Spheno}~2.2.3 \cite{Porod:2003um} online calculator \cite{
Allanach:2003jw, Belanger:2005jk,Kraml} with the SM input \cite{
PDBook}: $m_t = 174.2$ GeV, $m_b(m_b)^{\overline{\text{MS}}}=4.2$ GeV
and $\alpha_s^{\overline {\text{MS}}}(M_z) = 0.1176$. The relevant
masses of the SUSY particles are shown in Table \ref{SUSYparam} in
GeV. We have only listed those masses which are kinematically
accessible in the direct slepton decays of the given spectrum. In
these scenarios the lightest neutralino, $\tilde{\chi}_1^0$, is the
LSP.  Possible decay modes via the $\lam'_ {ijk}L_iQ_j\bar D_k$
interaction are \cite{Butterworth:1992tc,Dreiner:1994tj,Baltz:1997gd,
Richardson:2000nt}:
\begin{eqnarray}
\tilde{\chi}_1^0 \overset{\lambda'}{\longrightarrow}
\left\{ \begin{array}{cc}
\ell_i^+ \overline{u}_j d_k   & \\[1mm]
\ell_i^- u_j \overline{d}_k   & 
\end{array}
\right.
,\quad 
\tilde{\chi}_1^0 \overset{\lambda'}{\longrightarrow}
\left\{ \begin{array}{cc}
\overline{\nu}_i \overline{d}_j d_k & \\[1mm]
\nu_i d_j \overline{d}_k  &  
\end{array}
\right. . \quad
\label{neutralino_decay}
\end{eqnarray}
The heavier neutralinos $\tilde{\chi}_{2,3,4}^0$ and the charginos $
\tilde{\chi}_{1,2}^+$ may also decay via $\lambda'_{ijk}$ but these
channels are strongly suppressed compared to the gauge coupling decay
modes into the LSP. For example, the branching ratio ($BR$) for the
decay $\tilde{\chi}_{2}^0\overset{\lam'}{\longrightarrow}\mu^+{\bar u}
d$ with $\lambda'_{211}=0.01$ is less than $2\times 10^{-6}$ in the
five SPS scenarios, where the slepton can decay into an on-shell
$\tilde{\chi}_{2}^0$.

\begin{table}
\begin{ruledtabular}
\rule[0cm]{1.0\linewidth}{.5pt}\\[-1ex]
\begin{align*}
1. \qquad \tilde{\mu}^+ \rightarrow \mu^+\, &\tilde{\chi}_1^0 \\
        & \overset{\lam'}{\hookrightarrow}  \mu^+\,\bar{u}\,d \\[2ex]
2. \qquad \tilde{\mu}^+ \rightarrow \mu^+\, & \tilde{\chi}_{2,3,4}^0 \\
        & \hookrightarrow  Z^0 \, \neut{1} \\
                & \qquad\qquad \overset{\lam'}{\hookrightarrow}  \mu^+\, 
\bar{u}\, d \\
        & \hookrightarrow  \tilde{\mu}^-_R \, \mu^+ \\[-2ex] \\
3.  \qquad \tilde{\mu}^+ \rightarrow \bar{\nu}_{\mu}\, & \tilde{\chi}_{1,2}^+ 
\\
        & \hookrightarrow  W^+ \, \neut{1} \\
                & \qquad\qquad \overset{\lam'}{\hookrightarrow} (\nu_{\mu} 
\mu^+)\,(\mu^+ \bar{u}\, d) \\
4. \qquad \tilde{\nu}_{\mu}^* \rightarrow \mu^+\, &\tilde{\chi}_{1,2}^- \\
        & \hookrightarrow  W^- \, \neut{1} \\
                & \qquad\qquad \overset{\lam'}{\hookrightarrow}  \mu^+\, 
\bar{u} \,d 
\end{align*}
\rule{1.0\linewidth}{.5pt}
\caption{\label{tab_dimuon}
        Relevant decays of resonantly produced single smuons 
                $\tilde{\mu}$ and smuon sneutrinos $\tilde{\nu}_\mu$ 
                into like-sign muon pairs of positive charge.}
            \end{ruledtabular}
\end{table}

From the multiple decay modes we concentrate on those which lead to
final states with two like-sign leptons; a signature with low
background \cite{Baer:1989hr,Dreiner:1993ba,Barnett:1993ea,
  Guchait:1994zk,Dreiner:2000vf}. Here we assume $\lam'_{211}\not=0$,
since dimuon final states are of great interest for experimental
analyses \cite{Abe:1998gu,Affolder:2001nu,Acosta:2004kx,
  Abazov:2006ii}. For our numerical studies, we neglect decay modes
with branching ratios $BR < 0.5\%$ and where there are less than $0.1$
events for the given luminosity. The relevant decay chains leading to
a a like-sign muon pair are shown in Table \ref{tab_dimuon}.

There are two distinct baryon triality scenarios. If $\lam'_{211}$ is
comparable in size to the gauge coupling, the $\tilde{\mu}$ and
$\tilde{\nu}_\mu$ will have a significant decay width through the
$L_2Q_1\bar D_1$ operator. In contrast, if $\lam^\prime_{211}$ is much
smaller than the gauge couplings, the $\tilde{\mu}$ and $\tilde{\nu}_
\mu$ decay through $\mathrm{P}_6$-conserving channels. For example,
the $BR(\tilde{\mu}^+\overset{\lam'}{\longrightarrow} u {\bar d})$ is
less than $1.1\%$ for $\lam'_{211}=0.01$ in all six scenarios.  Baryon
triality then manifests itself mainly through the LSP decays.  Note
that the $BR$s of the LSP at tree level are independent of $\lam'_{211
}$. For the event rates, we choose $\lam'_{211}=0.05$ and $\lam'_{211
}=0.01$ in the following.

In Table \ref{dimuon_events}, we give the expected number of like-sign
dimuon events ($\mu^\pm\mu^\pm$) for the LHC (Tevatron) with an
integrated luminosity of $10{\rm fb}^{-1}$ ($1{\rm fb}^{-1}$).  We use
the NLO cross sections including QCD and SUSY-QCD corrections with the
tri-linear SUSY breaking parameter $A$ set to zero.  Apart from the
$\tilde{\mu}$, $\tilde{\nu}_{\mu}$ and LSP decays, we use the $BR$s
obtained by {\sc Spheno}~2.2.3 \cite{Porod:2003um}.

\begin{table*}[tb]
\begin{ruledtabular}
\begin{tabular}{l||cc|cc|cc|cc|cc|cc}
\multicolumn{13}{c}{$\lam'_{211}=0.05$} \\
\hline
$X$        & \multicolumn{2}{c|}{1a}  & \multicolumn{2}{c|}{1b} & 
\multicolumn{2}{c|}{2} & \multicolumn{2}{c|}{3} & \multicolumn{2}{c|}{4} 
& \multicolumn{2}{c}{5} \\
\hline
\hline
$\bar{u} d$  & 50246 & (647) & 15957 & (148) & 11.0 & (-) & 34972 & (373) 
& 1416 & (9.0) & 32126 & (370) \\
\hline
$\tilde{\mu}^-_R $   & \: 2997 & (38.6) & - & (-) & - & (-) & - & (-) & - 
& (-) & - & (-) \\
\hline
$Z^0 \bar{u} d$   & - & (-) & - & (-) & 39.5 & (-) & - & (-) & 3289 
& (20.8) & - & (-) \\
\hline
$\bar{\nu}_{\mu} \nu_{\mu} \bar{u} d$  & - & (-) & - & (-) & 7.6 & (-) 
& - & (-) & 613 & (3.9) & - & (-)\\
\hline
$W^- \bar{u} d $   & - & (-) & - & (-) & 52.1 & (-) & - & (-) & 5060 & (17.8) 
& 1752 & (13.8) \\
\hline
\hline
\multicolumn{13}{c}{$\lam'_{211}=0.01$} \\
\hline
$X$ & \multicolumn{2}{c|}{1a}  & \multicolumn{2}{c|}{1b} & \multicolumn{2}
{c|}{2} & \multicolumn{2}{c|}{3} & \multicolumn{2}{c|}{4} & \multicolumn{2}
{c}{5} \\
\hline
\hline
$\bar{u} d$  & 2279 & (29.3) & 723 & (6.7) & 0.4 & (-) & 1765 & (18.8) & 
57.8 & (0.4) & 1442 & (16.6) \\
\hline
$\tilde{\mu}^-_R $   & 136 & (1.7) & - & (-) & - & (-) & - & (-) &  - & (-) 
& - & (-) \\
\hline
$Z^0 \bar{u} d$   & - & (-) & - & (-) & 1.6 & (-) & - & (-) & 134 & (0.8) 
& - & (-) \\
\hline
$\bar{\nu}_{\mu} \nu_{\mu} \bar{u} d$   & - & (-) & - & (-) & 0.3 & (-) & - & 
(-) & 25.0 & (0.2) & - & (-) \\
\hline
$W^-\bar{u} d $   & - & (-) & - & (-) & 2.1 & (-) & - & (-) & 206 & (0.7) & 
80.1 & (0.6) \\
\end{tabular}
\caption{\label{dimuon_events} Number of like-sign dimuon events,
  $\mu^\pm\mu^\pm + X$, from cascade decays of $\tilde{\mu}$ and
  $\tilde{\nu}_\mu$ at the LHC (Tevatron) with an integrated
  luminosity of $10\,{\rm fb}^{-1}$ ($1\,{\rm fb}^{-1}$) for the six
  mSUGRA SPS scenarios. The expected number of events include also the
  charge conjugated final states.  Events which are kinematically
  forbidden or with $BR < 0.5\%$ or where there are less then $0.1$
  events are denoted by a dash.}
\end{ruledtabular}
\end{table*}

If decays into the heavier neutralinos $\tilde{\chi}^0_{3,4}$ and
chargino $\tilde{\chi}^+_{2}$ are kinematically forbidden (SPS1a, 1b,
3, 5), the $\tilde{\mu}$ decays dominantly into two like-sign muons
associated with two jets; contributions from the $\tilde{\nu}_\mu$
decay are negligible. Here all final state particles can be detected
and the slepton and gaugino masses can be reconstructed.  If the
$\tilde{\nu}_\mu$ and $\tilde{\mu}$ are heavier than all neutralinos
and charginos (SPS2, 4), then the $\tilde{\nu}_\mu$ decay gives the
main contribution and also decays with an on-shell $Z^0$ in the final
state are kinematically accessible.

In the case of $\tilde{e}$ and $\tilde{\nu}_e$ production, i.e.
$\lam'_{111}\not=0$, one obtains the same number of events for a given
value of $\lam'$, however the bounds on $\lam'_{111}$ are much
stronger \cite{Barbier:2004ez}.  If $\lam'_{311}\not=0$ there are more
possible decay modes, because left-right mixing in the stau sector can
not be neglected.  Both mass eigenstates $\tilde{\tau}_1$ and $\tilde
\tau_2$ can be produced via the $L_3Q_1\bar D_1$ operator and the heavier
mass eigenstate can decay into the lighter one via $\tilde{\tau}_2
\rightarrow \tilde{\tau}_1 Z^0$ and $\tilde{\tau}_2 \rightarrow \tilde
{\tau}_1 h^0$, if kinematically allowed. This results in more decay
chains compared to those for $\tilde{\mu}$ and $\tilde{\nu}_{\mu}$.
In addition, the charge of the $\tau$-jets has to be reconstructed.

At the LHC, the beam energy will be high enough to produce single
sleptons at a high rate and an excess of like-sign lepton pairs would
be a hint of baryon-triality supersymmetry. Since the event rates
scale with $|\lam'_{ijk}|^2$ for $\lam'_{ijk}$ much smaller than the
gauge couplings, we can roughly estimate that the magnitude of
$\lam'_{211}$ can be tested down to $10^{-2}-10^{-4}$, depending on
the scenario.


\section{Conclusion}
\label{conclusion}

\cleqn 

We have argued that in preparation for the LHC, the baryon-triality
interactions must be considered in detail. We have computed the full
NLO corrections to resonant slepton production at hadron colliders. We
have found significant changes in the total and differential cross
sections. In particular, we have computed the SUSY-QCD corrections for
the first time and found that the effects can be large in specific
regions of parameter space, modifying the QCD prediction by up to
35\%. The factorization scale dependence of the cross section 
at the LHC (Tevatron) is also significantly reduced to about $\pm$5\%
($\pm$10\%) at NLO.  We have used our
computation to determine the running of the $\lam'$-coupling without
and with the decoupling of heavy SUSY particles.

Next, we considered the slepton tranverse momentum distribution and
resummed the final state gluons. Including these results we have
performed a detailed comparison with the {\sc Herwig} and {\sc
  Susygen}/{\sc Pythia} Monte Carlo generators. Significant
discrepancies remain, similar to the case of SM $Z$-boson production.
This needs to be remedied. Finally including the slepton decay
branching ratios we determine the like-sign dimuon event rates at the
LHC and Tevatron for various SUSY spectra. This should enable a
detailed investigation by the experimental groups.

\begin{acknowledgments}
The authors want to thank Howard E.\ Haber, Tilman Plehn and Michael
Spira for helpful discussions concerning the running of $\lambda'$.
We want to thank Peter Richardson for helping us to interpret the
Monte Carlo spectra, Christian Autermann for helping us with {\sc
Susygen}, Nicolas M{\"o}ser for generating transverse momentum
distributions of the $Z$ boson, and Thomas Hebbeker for discussions on
experimental aspects of resonant slepton searches. We also thank Olaf
Kittel for reading parts of the manuscript.
\end{acknowledgments}

\appendix

\section{High-$p_T$ Slepton Distribution at 
  $\mathcal{O}(\alpha_s)$}
\label{pertpT}

The transverse momentum spectrum in the high $p_T$ region, i.e. $p_T
\gsim \tilde{m}$, can be calculated perturbatively.  According to
\cite{Altarelli:1984pt,Kauffman:1991jt} the perturbative hadron level
cross section for the parton level process
\begin{eqnarray}
a(p_1)+b(p_2) \rightarrow \tilde L(q) + X\,,
\end{eqnarray}
can be written as:
\begin{align}
\begin{split}
\frac{d\sigma^\text{pert}}{dp_Tdy} &= \\
        \frac{2p_T}{S}  \Big[& \int_{x_A^*}^1 \frac{dx_1}{x_1-x_1^+} 
                f_{aA}(x_1)f_{bB}(x_2^*)\hat{s}\frac{d\hat{\sigma}^{ab}}
{d\hat{t}}(x_2=x_2^*)  \\ 
        +& \int_{x_B^*}^1 \frac{dx_2}{x_2-x_2^+} 
f_{aA}(x_1^*)f_{bB}(x_2)\hat{s}\frac{d\hat{\sigma}^{ab}}{d\hat{t}}(x_1=x_1^*)
\Big ] \, .
\label{pert}
\end{split}
\end{align}
Here, parton $a$ with momentum $p_1=x_1P_1$, and parton $b$ with
momentum $p_2=x_2P_2$ produce a slepton (sneutrino) ($\tilde L$) with
momentum $q$ and mass $\tilde{m}$ and an additional particle, $X$.
Parton $a$ originates from hadron $A$ with momentum $P_1$ and parton
$b$ originates from hadron $B$ with momentum $P_2$. $f_{aA}(x_1)$
denotes the parton density function for the parton $a$ in hadron $A$
with momentum fraction $x_1$. The other hadron level variables as
functions of $p_T$ are
\begin{align}
S &= (P_1+P_2)^2, \nonumber \\ 
\sqrt{\tau^+} &= \sqrt{p_T^2/S} + \sqrt{(\tilde{m}^2+p_T^2)/S}, \nonumber\\
x_{A/B}^* &= e^{\pm y} \sqrt{\tau^+}, \qquad \quad 
x_{1/2}^0 = e^{\pm y}\sqrt{\tilde{m}^2/S}, \nonumber \\
x_{1/2}^+ &= e^{\pm y}\sqrt{(\tilde{m}^2+p_T^2)/S} , \nonumber\\
x_{1/2}^*&=(x_{2/1}x_{1/2}^+-\tilde{m}^2/S)/(x_{2/1}-x_{2/1}^+)\,.
\end{align}
Here $y$ is the slepton rapidity and $x^0_{1/2}$ are the parton
momentum fractions in the limit $p_T/\tilde{m}\rightarrow0$. The partonic
variables are given by
\begin{align}
\hat{s} &= x_1x_2S, \nonumber \\
\hat{t} &= \tilde{m}^2 \left( 1 - \frac{x_1}{x_1^0}\sqrt{1+\frac{p_T^2}
{\tilde{m}^2}} \right) \, , \nonumber \\
\hat{u} &= \tilde{m}^2 \left( 1 - \frac{x_2}{x_2^0}\sqrt{1+\frac{p_T^2}
{\tilde{m}^2}} \right) \, .
\label{mandelstam}
\end{align}
The transverse momentum spectrum of the slepton (sneutrino) is at $O(
\alpha_s)$ due to the gluon radiation and Compton-like processes. For
gluon radiation, the relevant parton level processes are
\begin{align}
&q(p_1)+\overline{q}'(p_2) \rightarrow \tilde L(q) + g , \nonumber\\
&\overline{q}'(p_1)+q(p_2) \rightarrow \tilde L(q) + g \, ,
\end{align} 
where $q$ and $q'$ are not necessarily the same flavour. The resulting
differential partonic cross section is given by
\begin{align}
\hat{s}\frac{d\hat{\sigma}^{q\overline{q}'}}{d\hat{t}} &=
\hat{s}\frac{d\hat{\sigma}^{\overline{q}'q}}{d\hat{t}}  = 
\frac{\alpha_s\lam'^2}{18\hat{s}}
\left[ \frac{\hat{u}}{\hat{t}} + \frac{\hat{t}}{\hat{u}} + 2\left( 
\frac{\tilde{m}^2\hat{s}}{\hat{u}\hat{t}}+1 \right) \right]\, ,
\label{qqbar}
\end{align}
and $\hat s,\, \hat t,\, \hat u$ are the parton level Mandelstam
variables (\ref{mandelstam}). 
The parton level Compton-like processes with 
$a=q,\overline{q}'$, $b=q',\overline{q}$ 
and $g$ denoting a gluon are:
\begin{align}
&a(p_1) + g(p_2) \rightarrow \tilde L(q) + b\, , \nonumber\\
&g(p_1) + a(p_2) \rightarrow \tilde L(q) + b \, .
\end{align} 
We obtain the differential partonic cross sections:
\begin{align}
\hat{s}\frac{d\hat{\sigma}^{ag}}{d\hat{t}} &= -\frac{\alpha_s\lam'^2}
{48\hat{s}} \left[ \frac{\hat{s}}{\hat{t}} + \frac{\hat{t}}{\hat{s}} + 
2\left( \frac{\tilde{m}^2\hat{u}}{\hat{s}\hat{t}}+1 \right) \right]\, ,
 \nonumber \\
\hat{s}\frac{d\hat{\sigma}^{ga}}{d\hat{t}} &= -\frac{\alpha_s\lam
'^2}{48\hat{s}} \left[ \frac{\hat{s}}{\hat{u}} + \frac{\hat{u}}{\hat
{s}} + 2\left( \frac{\tilde{m}^2\hat{t}}{\hat{s}\hat{u}}+1 \right) \right] \, .
\label{ga}
\end{align}
These can then be inserted in Eq.~(\ref{pert}) together with the parton
density functions to obtain the corresponding hadronic differential
cross sections.



\bibliographystyle{h-physrev}

\bibliography{references}

\begin{thebibliography}{10}

\bibitem{Wess:1974tw}
J.~Wess and B.~Zumino,
\newblock Nucl. Phys. {\bf B70}, 39 (1974).

\bibitem{Sakai:1981gr}
N.~Sakai,
\newblock Zeit. Phys. {\bf C11}, 153 (1981).

\bibitem{Witten:1981nf}
E.~Witten,
\newblock Nucl. Phys. {\bf B188}, 513 (1981).

\bibitem{Veltman:1980mj}
M.~J.~G. Veltman,
\newblock Acta Phys. Polon. {\bf B12}, 437 (1981).

\bibitem{Kaul:1981hi}
R.~K. Kaul and P.~Majumdar,
\newblock Nucl. Phys. {\bf B199}, 36 (1982).

\bibitem{Glashow:1961tr}
S.~L. Glashow,
\newblock Nucl. Phys. {\bf 22}, 579 (1961).

\bibitem{Weinberg:1967tq}
S.~Weinberg,
\newblock Phys. Rev. Lett. {\bf 19}, 1264 (1967).

\bibitem{Nilles:1983ge}
H.~P. Nilles,
\newblock Phys. Rept. {\bf 110}, 1 (1984).

\bibitem{Martin:1997ns}
S.~P. Martin,
\newblock (1997), hep-ph/9709356.

\bibitem{Drees:1996ca}
M.~Drees,
\newblock (1996), hep-ph/9611409.

\bibitem{Sakai:1981pk}
N.~Sakai and T.~Yanagida,
\newblock Nucl. Phys. {\bf B197}, 533 (1982).

\bibitem{Weinberg:1981wj}
S.~Weinberg,
\newblock Phys. Rev. {\bf D26}, 287 (1982).

\bibitem{Dreiner:1997uz}
H.~K. Dreiner,
\newblock (1997), hep-ph/9707435.

\bibitem{Smirnov:1996bg}
A.~Y. Smirnov and F.~Vissani,
\newblock Phys. Lett. {\bf B380}, 317 (1996), hep-ph/9601387.

\bibitem{Shiozawa:1998si}
Super-Kamiokande, M.~e.~a. Shiozawa,
\newblock Phys. Rev. Lett. {\bf 81}, 3319 (1998), hep-ex/9806014.

\bibitem{Dreiner:2005rd}
H.~Dreiner, C.~Luhn, and M.~Thormeier,
\newblock Phys. Rev. {\bf D73}, 075007 (2006), hep-ph/0512163.

\bibitem{Ibanez:1991hv}
L.~E. Ib\'a\~nez and G.~G. Ross,
\newblock Phys. Lett. {\bf B260}, 291 (1991).

\bibitem{Ibanez:1991pr}
L.~E. Ib\'a\~nez and G.~G. Ross,
\newblock Nucl. Phys. {\bf B368}, 3 (1992).

\bibitem{Dimopoulos:1981dw}
S.~Dimopoulos, S.~Raby, and F.~Wilczek,
\newblock Phys. Lett. {\bf B112}, 133 (1982).

\bibitem{Barbier:2004ez}
R.~Barbier {\em et~al.},
\newblock Phys. Rept. {\bf 420}, 1 (2005), hep-ph/0406039.

\bibitem{Dreiner:1991pe}
H.~Dreiner and G.~G. Ross,
\newblock Nucl. Phys. {\bf B365}, 597 (1991).

\bibitem{Allanach:1999bf}
R parity Working Group, B.~Allanach {\em et~al.},
\newblock (1999), hep-ph/9906224.

\bibitem{Allanach:2003eb}
B.~C. Allanach, A.~Dedes, and H.~K. Dreiner,
\newblock Phys. Rev. {\bf D69}, 115002 (2004), hep-ph/0309196.

\bibitem{Allanach:2002nj}
B.~C. Allanach {\em et~al.},
\newblock Eur. Phys. J. {\bf C25}, 113 (2002), hep-ph/0202233.

\bibitem{Breitweg:1997ff}
ZEUS, J.~Breitweg {\em et~al.},
\newblock Z. Phys. {\bf C74}, 207 (1997), hep-ex/9702015.

\bibitem{Adloff:1997fg}
H1, C.~Adloff {\em et~al.},
\newblock Z. Phys. {\bf C74}, 191 (1997), hep-ex/9702012.

\bibitem{Dreiner:1997cd}
H.~K. Dreiner and P.~Morawitz,
\newblock Nucl. Phys. {\bf B503}, 55 (1997), hep-ph/9703279.

\bibitem{Altarelli:1997ce}
G.~Altarelli, J.~R. Ellis, G.~F. Giudice, S.~Lola, and M.~L. Mangano,
\newblock Nucl. Phys. {\bf B506}, 3 (1997), hep-ph/9703276.

\bibitem{Kalinowski:1997fk}
J.~Kalinowski, R.~Ruckl, H.~Spiesberger, and P.~M. Zerwas,
\newblock Z. Phys. {\bf C74}, 595 (1997), hep-ph/9703288.

\bibitem{Plehn:1997az}
T.~Plehn, H.~Spiesberger, M.~Spira, and P.~M. Zerwas,
\newblock Z. Phys. {\bf C74}, 611 (1997), hep-ph/9703433.

\bibitem{Kramer:1997hh}
M.~Kramer, T.~Plehn, M.~Spira, and P.~M. Zerwas,
\newblock Phys. Rev. Lett. {\bf 79}, 341 (1997), hep-ph/9704322.

\bibitem{Dimopoulos:1988jw}
S.~Dimopoulos and L.~J. Hall,
\newblock Phys. Lett. {\bf B207}, 210 (1988).

\bibitem{Dimopoulos:1988fr}
S.~Dimopoulos, R.~Esmailzadeh, L.~J. Hall, and G.~D. Starkman,
\newblock Phys. Rev. {\bf D41}, 2099 (1990).

\bibitem{Dreiner:1998gz}
H.~K. Dreiner, P.~Richardson, and M.~H. Seymour,
\newblock (1998), hep-ph/9903419.

\bibitem{Dreiner:2000qf}
H.~K. Dreiner, P.~Richardson, and M.~H. Seymour,
\newblock (2000), hep-ph/0001224.

\bibitem{Dreiner:2000vf}
H.~K. Dreiner, P.~Richardson, and M.~H. Seymour,
\newblock Phys. Rev. {\bf D63}, 055008 (2001), hep-ph/0007228.

\bibitem{Marchesini:1991ch}
G.~Marchesini {\em et~al.},
\newblock Comput. Phys. Commun. {\bf 67}, 465 (1992).

\bibitem{Corcella:2000bw}
G.~Corcella {\em et~al.},
\newblock JHEP {\bf 01}, 010 (2001), hep-ph/0011363.

\bibitem{Corcella:2002jc}
G.~Corcella {\em et~al.},
\newblock (2002), hep-ph/0210213.

\bibitem{Moretti:2002eu}
S.~Moretti, K.~Odagiri, P.~Richardson, M.~H. Seymour, and B.~R. Webber,
\newblock JHEP {\bf 04}, 028 (2002), hep-ph/0204123.

\bibitem{Choudhury:2002au}
D.~Choudhury, S.~Majhi, and V.~Ravindran,
\newblock Nucl. Phys. {\bf B660}, 343 (2003), hep-ph/0207247.

\bibitem{Yang:2005ts}
L.~L. Yang, C.~S. Li, J.~J. Liu, and Q.~Li,
\newblock Phys. Rev. {\bf D72}, 074026 (2005), hep-ph/0507331.

\bibitem{Nilles:1982dy}
H.~P. Nilles, M.~Srednicki, and D.~Wyler,
\newblock Phys. Lett. {\bf B120}, 346 (1983).

\bibitem{Ghodbane:1999va}
N.~Ghodbane,
\newblock (1999), hep-ph/9909499.

\bibitem{Plehn:2000be}
T.~Plehn,
\newblock Phys. Lett. {\bf B488}, 359 (2000), hep-ph/0006182.

\bibitem{Butterworth:1992tc}
J.~Butterworth and H.~K. Dreiner,
\newblock Nucl. Phys. {\bf B397}, 3 (1993), hep-ph/9211204.

\bibitem{Kon:1991ad}
T.~Kon and T.~Kobayashi,
\newblock Phys. Lett. {\bf B270}, 81 (1991).

\bibitem{Dreiner:1995ij}
H.~K. Dreiner and S.~Lola,
\newblock Prepared for Physics with e+ e- Linear Colliders (The European
  Working Groups 4 Feb - 1 Sep 1995: Session 3), Hamburg, Germany, 30 Aug - 1
  Sep 1995.

\bibitem{Erler:1996ww}
J.~Erler, J.~L. Feng, and N.~Polonsky,
\newblock Phys. Rev. Lett. {\bf 78}, 3063 (1997), hep-ph/9612397.

\bibitem{Allanach:1999ic}
B.~C. Allanach, A.~Dedes, and H.~K. Dreiner,
\newblock Phys. Rev. {\bf D60}, 075014 (1999), hep-ph/9906209.

\bibitem{Chemtob:2004xr}
M.~Chemtob,
\newblock Prog. Part. Nucl. Phys. {\bf 54}, 71 (2005), hep-ph/0406029.

\bibitem{Haber:1984rc}
H.~E. Haber and G.~L. Kane,
\newblock Phys. Rept. {\bf 117}, 75 (1985).

\bibitem{Rosiek:1995kg}
J.~Rosiek,
\newblock Phys. Rev {\bf D41}, 3464 (1990),
\newblock with Errata hep-ph/9511250.

\bibitem{Soni:1983rm}
S.~K. Soni and H.~A. Weldon,
\newblock Phys. Lett. {\bf B126}, 215 (1983).

\bibitem{Denner:1991kt}
A.~Denner,
\newblock Fortschr. Phys. {\bf 41}, 307 (1993).

\bibitem{Martin:1993zk}
S.~P. Martin and M.~T. Vaughn,
\newblock Phys. Rev. {\bf D50}, 2282 (1994), hep-ph/9311340.

\bibitem{Pumplin:2005rh}
J.~Pumplin, A.~Belyaev, J.~Huston, D.~Stump, and W.~K. Tung,
\newblock JHEP {\bf 02}, 032 (2006), hep-ph/0512167.

\bibitem{Martin:2002dr}
A.~D. Martin, R.~G. Roberts, W.~J. Stirling, and R.~S. Thorne,
\newblock Phys. Lett. {\bf B531}, 216 (2002), hep-ph/0201127.

\bibitem{Martin:2004ir}
A.~D. Martin, R.~G. Roberts, W.~J. Stirling, and R.~S. Thorne,
\newblock Phys. Lett. {\bf B604}, 61 (2004), hep-ph/0410230.

\bibitem{Collins:1981uk}
J.~C. Collins and D.~E. Soper,
\newblock Nucl. Phys. {\bf B193}, 381 (1981).

\bibitem{Collins:1981va}
J.~C. Collins and D.~E. Soper,
\newblock Nucl. Phys. {\bf B197}, 446 (1982).

\bibitem{Collins:1984kg}
J.~C. Collins, D.~E. Soper, and G.~Sterman,
\newblock Nucl. Phys. {\bf B250}, 199 (1985).

\bibitem{Han:1991sa}
T.~Han, R.~Meng, and J.~Ohnemus,
\newblock Nucl. Phys. {\bf B384}, 59 (1992).

\bibitem{Altarelli:1977zs}
G.~Altarelli and G.~Parisi,
\newblock Nucl. Phys. {\bf B126}, 298 (1977).

\bibitem{Parisi:1979se}
G.~Parisi and R.~Petronzio,
\newblock Nucl. Phys. {\bf B154}, 427 (1979).

\bibitem{Davies:1984sp}
C.~T.~H. Davies, B.~R. Webber, and W.~J. Stirling,
\newblock Nucl. Phys. {\bf B256}, 413 (1985).

\bibitem{Landry:2002ix}
F.~Landry, R.~Brock, P.~M. Nadolsky, and C.~P. Yuan,
\newblock Phys. Rev. {\bf D67}, 073016 (2003), hep-ph/0212159.

\bibitem{Qiu:2000hf}
J.-w. Qiu and X.-f. Zhang,
\newblock Phys. Rev. {\bf D63}, 114011 (2001), hep-ph/0012348.

\bibitem{Hinchliffe:1988ap}
I.~Hinchliffe and S.~F. Novaes,
\newblock Phys. Rev. {\bf D38}, 3475 (1988).

\bibitem{Kauffman:1991jt}
R.~P. Kauffman,
\newblock Phys. Rev. {\bf D44}, 1415 (1991).

\bibitem{Sjostrand:2001yu}
T.~Sjostrand, L.~Lonnblad, and S.~Mrenna,
\newblock (2001), hep-ph/0108264.

\bibitem{Mrenna:1996hu}
S.~Mrenna,
\newblock Comput. Phys. Commun. {\bf 101}, 232 (1997), hep-ph/9609360.

\bibitem{Peter}
P.~Richardson,
\newblock private communication.

\bibitem{Porod:2003um}
W.~Porod,
\newblock Comput. Phys. Commun. {\bf 153}, 275 (2003), hep-ph/0301101.

\bibitem{Allanach:2003jw}
B.~C. Allanach, S.~Kraml, and W.~Porod,
\newblock JHEP {\bf 03}, 016 (2003), hep-ph/0302102.

\bibitem{Belanger:2005jk}
G.~Belanger, S.~Kraml, and A.~Pukhov,
\newblock Phys. Rev. {\bf D72}, 015003 (2005), hep-ph/0502079.

\bibitem{Kraml}
S.~Kraml,
\newblock http://cern.ch/kraml/comparison/.

\bibitem{PDBook}
W.-M. Yao {\em et~al.},
\newblock Journal of Physics G {\bf 33}, {1+} (2006),
\newblock http://pdg.lbl.gov.

\bibitem{Dreiner:1994tj}
H.~K. Dreiner and P.~Morawitz,
\newblock Nucl. Phys. {\bf B428}, 31 (1994), hep-ph/9405253.

\bibitem{Baltz:1997gd}
E.~A. Baltz and P.~Gondolo,
\newblock Phys. Rev. {\bf D57}, 2969 (1998), hep-ph/9709445.

\bibitem{Richardson:2000nt}
P.~Richardson,
\newblock (2000), hep-ph/0101105.

\bibitem{Baer:1989hr}
H.~Baer, X.~Tata, and J.~Woodside,
\newblock Phys. Rev. {\bf D41}, 906 (1990).

\bibitem{Dreiner:1993ba}
H.~K. Dreiner, M.~Guchait, and D.~P. Roy,
\newblock Phys. Rev. {\bf D49}, 3270 (1994), hep-ph/9310291.

\bibitem{Barnett:1993ea}
R.~M. Barnett, J.~F. Gunion, and H.~E. Haber,
\newblock Phys. Lett. {\bf B315}, 349 (1993), hep-ph/9306204.

\bibitem{Guchait:1994zk}
M.~Guchait and D.~P. Roy,
\newblock Phys. Rev. {\bf D52}, 133 (1995), hep-ph/9412329.

\bibitem{Abe:1998gu}
CDF, F.~Abe {\em et~al.},
\newblock Phys. Rev. Lett. {\bf 83}, 2133 (1999), hep-ex/9908063.

\bibitem{Affolder:2001nu}
CDF, A.~A. Affolder {\em et~al.},
\newblock Phys. Rev. Lett. {\bf 87}, 251803 (2001), hep-ex/0106061.

\bibitem{Acosta:2004kx}
CDF, D.~Acosta {\em et~al.},
\newblock Phys. Rev. Lett. {\bf 93}, 061802 (2004), hep-ex/0405063.

\bibitem{Abazov:2006ii}
D0, V.~M. Abazov {\em et~al.},
\newblock (2006), hep-ex/0605010.

\bibitem{Altarelli:1984pt}
G.~Altarelli, R.~K. Ellis, M.~Greco, and G.~Martinelli,
\newblock Nucl. Phys. {\bf B246}, 12 (1984).

\bibitem{Alves:2002tj}
A.~Alves, O.~Eboli, and T.~Plehn,
\newblock Phys. Lett. {\bf B558}, 165 (2003), hep-ph/0211441.

\end{thebibliography}

\end{document}